\documentclass[aps,pra,twocolumn,groupedaddress]{revtex4-2}
\usepackage{amsmath}
\usepackage{xcolor}
\usepackage{commath}
\usepackage{graphicx}
\usepackage{hyperref}
\usepackage{lmodern}
\usepackage[T1]{fontenc}
\DeclareMathOperator{\tr}{tr}

\newcommand{\delnospace}[1]{\mathopen{}\del{#1}}
\begin{document}

\title{Spin amplification in realistic systems}

\author{Ivan Iakoupov$^{1,2}$}
\email{ivan.iakoupov@oist.jp}
\author{Victor M. Bastidas$^{2}$}
\author{Yuichiro Matsuzaki$^{2,3}$}
\author{Shiro Saito$^{2}$}
\author{William J. Munro$^{1,2}$}
\affiliation{$^{1}$Quantum Engineering and Design Unit, Okinawa Institute of 
Science and Technology Graduate University, 1919-1 Tancha, Onna-son, Okinawa 
904-0495, Japan\\
$^{2}$Basic Research Laboratories, NTT, Inc., 3-1 Morinosato-Wakamiya, Atsugi, Kanagawa 243-0198, Japan\\
$^{3}$Department of Electrical, Electronic, and Communication Engineering, Faculty of Science and Engineering, Chuo University, 1-13-27 Kasuga, Bunkyo-ku, Tokyo 112-8551, Japan
}

\date{\today}

\begin{abstract}
Spin amplification is the process that ideally increases the number of excited 
spins when one of them is excited initially. We show that by applying 
optimal control techniques to design classical drive pulse shapes, spin 
amplification can be achieved in a previously unexplored fast regime, with 
amplification times comparable to the intrinsic interaction timescale.
This is an order of magnitude faster than the previous protocols and makes spin 
amplification possible even with significant decoherence and inhomogeneity in the spin 
system. The initial spin excitation can be delocalized over the entire 
ensemble, which is a more typical situation when a photon is collectively 
absorbed by the spins. We focus on the superconducting persistent-current 
artificial atoms and the Rydberg atoms as spins.
\end{abstract}


\maketitle
\section{Introduction}
Control of ever larger quantum systems is essential for quantum 
simulation~\cite{arute_science20,ebadi_nature21,scholl_nature21,kim_nature23}, 
quantum sensing~\cite{bornet_nature23,eckner_nature23,franke_nature23,hines_prl23} and 
error-corrected quantum 
computing~\cite{chen_prl22,zhao_prl22,acharya_nature23}. Such quantum systems
can be viewed as collections of spins encoded in real or artificial atoms. 
Instead of aiming to deliver a separate driving field for each spin, experimental 
complexity could be decreased by using a global drive for the entire spin 
ensemble~\cite{dreves_prl83,benjamin_pra00,benjamin_prl01,mintert_prl01,fitzsimons_prl06,lee_pra05,furman_prb06,close_prl11,glaetzle_prx17,cesa_prl23,patomaki_npjqi24,nill_prl24}.
Addressing the individual spins or spin groups can then be done by frequency 
or polarization selectivity and engineering of the spin-spin interactions.

\begin{figure}[b]
\begin{center}
\includegraphics{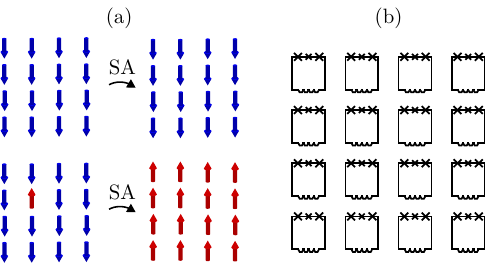}
\end{center}
\caption{(a) The idealized spin amplification (SA) operation. If all spins are 
in the ground state initially, there is no change (top). If one spin is excited 
initially, all spins become excited (bottom). We consider an equal superposition of all 
the single excitation states instead of a 
particular excited spin position as shown in the figure. (b) One of the considered 
experimental platforms: a 2D array of persistent-current artificial 
atoms~\cite{orlando_prb99} as the spin ensemble. The artificial atoms consist of superconducting 
loops interrupted by $3$ Josephson junctions (crosses) where one of the 
junctions is smaller than the other two. The superconducting 
loops have linear inductance (inductors at the bottom of the loops), and 
the interactions between the spins are due to the mutual inductance between the 
loops. The interactions are all-to-all in nature.
\label{fig_setup}}
\end{figure}

One of the operations that is useful for quantum sensing is
spin amplification~\cite{cappellaro_qcss06,yoshinaga_pra21,nill_prl24},
which can be implemented with a global drive~\cite{lee_pra05,furman_prb06,close_prl11}. 
The idealized version of this operation amplifies a single 
excited spin into many excited spins, while keeping the ensemble with zero 
excited spins unchanged [see Fig.~\ref{fig_setup}(a)]. This can be realized using the 
transverse-field Ising model with nearest-neighbor interactions. A 
continuous-wave weak classical field is applied such that it is off-resonant 
with the frequency of the spins. One excited spin shifts the frequency of the 
neighboring spins into resonance with the drive, making a domain wall 
propagate in such a way that the number of the excited spins is 
increased~\cite{lee_pra05,furman_prb06,close_prl11,coldea_science10,nill_prl24}.

The transverse-field Ising model can be implemented on general-purpose 
quantum computers~\cite{kim_nature23}, but this requires individual drives for
the spins. In contrast, ensembles of Rydberg atoms for instance, implement 
this model while utilizing a global 
drive~\cite{ebadi_nature21,scholl_nature21}. An implementation of 
the transverse-field Ising model can also be achieved by the ensembles of 
superconducting persistent-current artificial atoms~\cite{orlando_prb99,harris_prb09}. Compared to other 
superconducting artificial atoms with much better coherence 
properties~\cite{place_ncomms21,wang_npjqi22,somoroff_prl23,tuokkola_ncomms25}, the primary 
advantage of the persistent-current artificial atoms is the small size that 
permits fabrication of thousands of spins on a compact 
chip~\cite{kakuyanagi_prl16}. Their large anharmonicity makes it possible to treat them 
as two-level systems even with strong drives, although large anharmonicities 
could also be obtained with for instance fluxonium~\cite{somoroff_prl23} at the cost 
of a larger footprint. Due to being related to the persistent-current 
artificial atom, fluxonium can also have native 
$ZZ$-interactions~\cite{ma_prl24}. This allows a fluxonium ensemble to be another 
possible implementation of the transverse-field Ising model.

Realistic spin systems have imperfections, such as long-range interactions, 
finite lifetimes, and inhomogeneous 
broadening~\cite{kakuyanagi_prl16,ebadi_nature21,scholl_nature21,chang_prapplied22}.  
The position of the initial single excitation can be delocalized over the 
entire ensemble due to spins collectively absorbing a 
photon~\cite{lukin_rmp03,hammerer_rmp10}. Designing a protocol to amplify such 
delocalized states using imperfect spins is difficult to do manually. 
Therefore, we turn to optimal control~\cite{brif_njp10} to replace the weak 
continuous-wave drive field with a time-dependent one that can be much 
stronger at its maximum.

The rest of the article is organized as follows. Our theoretical model and 
parameters are described in Sec.~\ref{sec_setup} with the results presented in 
Sec.~\ref{sec_results}. The implications of the results are discussed in 
Sec.~\ref{sec_discussion_conclusion} with a conclusion also provided.

\section{Setup}\label{sec_setup}
We focus on the 2D layout of the spins (depicted schematically in 
Fig.~\ref{fig_setup}) due to it being easily transferable to the 
experiments~\cite{ebadi_nature21,scholl_nature21,kakuyanagi_prl16}. The spins 
are encoded either in the Rydberg atoms or in the persistent-current 
artificial atoms. The Hamiltonians for both systems are in the interaction 
picture with respect to the carrier frequency of the drive $\omega_\text{d}$ 
and use the rotating wave approximation.

The Hamiltonian for the ensemble of the Rydberg atoms is
\begin{gather}\label{H_long_range_ising_Rydberg}
H_\text{Ryd}=\hbar\sum_j\left(-\Delta\hat{n}_{j}
+\Omega\hat{\sigma}_{x,j}\right)
+\hbar\sum_{\substack{j,k\\j\neq k}} J_{z,jk} \hat{n}_{j}\hat{n}_{k},
\end{gather}
where $\hat{n}_j=|1\rangle\langle 1|_j=(1+\hat{\sigma}_{z,j})/2$ is the 
projector onto the excited state of the spin $j$ ($\hat{\sigma}_{z,j}$ is the 
Pauli-$Z$ operator), $\Delta=\omega_\text{d}-\omega_{01}$ is the detuning 
between the transition frequency $\omega_{01}$ and $\omega_\text{d}$, and 
$J_{z,jk}\propto r_{jk}^{-6}$ with $r_{jk}$ is the distance between the 
spins $j$ and $k$. The choices for the absent factor of~$2$ in the $\Omega$ 
term and double-counting of the $J_{z,jk}$ terms (by not restricting $j<k$ in 
the summation) are different from Refs.~\cite{ebadi_nature21,scholl_nature21}, 
so rescaling of the parameters is necessary. Defining $J_{z,\text{nn}}$ to be 
the $J_{z,jk}$ where $j$ and $k$ are nearest neighbors, we therefore set 
$J_{z,\text{nn}}/(2\pi)=1$~MHz, and maximum Rabi frequency
$\Omega/(2\pi)=3.5$~MHz~\cite{scholl_nature21}. Both the detuning $\Delta$ and 
the Rabi frequency $\Omega$ can be time-dependent to perform optimal control 
of the Rydberg atoms~\cite{omran_science19}. The atoms are assumed to be 
trapped in a square grid with the nearest neighbors $10$ $\mu$m 
apart~\cite{scholl_nature21}. The errors in positions have two components: the 
static trap position offset with the standard deviation of $100$ nm, and the 
position of the atom within the trap due to the finite temperature with 
standard deviations of $170$ nm in the plane of the array and $1$ $\mu$m out 
of the plane~\cite{scholl_nature21}. These errors translate into inhomogeneity 
of the couplings $J_{z,jk}$.

The Hamiltonian for the ensemble of persistent-current artificial atoms is
\begin{gather}\label{H_long_range_ising}
\begin{gathered}
H_\text{PCAA}=-\hbar\sum_j\left(\frac{\Delta_j}{2}\hat{\sigma}_{z,j}
+f_j\Omega\hat{\sigma}_{+,j}
+f_j\Omega^* \hat{\sigma}_{-,j}\right)\\
+\hbar\sum_{\substack{j,k\\j\neq k}} J_{z,jk} \hat{\sigma}_{z,j}\hat{\sigma}_{z,k}
+\hbar\sum_{\substack{j,k\\j\neq k}} J_{\pm,jk} \hat{\sigma}_{+,j}\hat{\sigma}_{-,k},
\end{gathered}
\end{gather}
where $\hat{\sigma}_{\pm,j}$ is the raising/lowering operator. In contrast to 
the Rydberg atoms, the transition frequencies $\omega_{01,j}$ and hence the 
detunings $\Delta_j$ are different for the different spins. They can be 
written $\Delta_j=\Delta_\text{avg}+\omega_{{01},\text{avg}}-\omega_{01,j}$, 
where $\omega_{01,\text{avg}}$ is the average frequency (with averaging both 
over the spins in the ensemble and inhomogeneity realizations of the 
ensemble), and $\Delta_\text{avg}=\omega_\text{d}-\omega_{{01},\text{avg}}$. 
The Rabi frequency $\Omega$ can be time-dependent and complex, with the real 
and imaginary parts corresponding to the $I$ and $Q$ signals.

\begin{figure}[t]
\begin{center}
\includegraphics{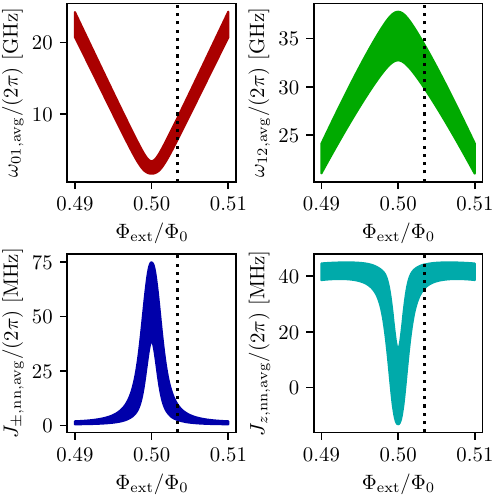}
\end{center}
\caption{The average parameters of the Hamiltonian~\eqref{H_long_range_ising} 
for the persistent-current artificial atoms~\cite{orlando_prb99} as spins: the 
transition frequencies between the ground and excited state 
$\omega_{01,\text{avg}}$, the transition frequencies between the first and 
second excited states $\omega_{12,\text{avg}}$, and the values of the 
nearest-neighbor couplings $J_{\pm,\text{nn},\text{avg}}$ and 
$J_{z,\text{nn},\text{avg}}$ ($J_{\pm,jk}$ and $J_{z,jk}$ with the maximal 
absolute values, respectively). Values for a range of bias fluxes 
$\Phi_\text{ext}/\Phi_0$ are shown, where $\Phi_0$ is the magnetic flux 
quantum. The curves are calculated by performing circuit quantization of the 
persistent-current artificial atoms with the Josephson junction area error of 
$2\%$ for $10^4$ different realizations of a square $4\times4$ ensemble. The 
difference in the shown values between using $3\times3$ and $4\times4$ 
ensembles for averaging is small. The thickness of the curves shows the 
standard deviation of the parameters. The chosen 
$\Phi_\text{ext}/\Phi_0=0.5034$ is shown by the black dotted vertical line. 
The parameters of the persistent-current artificial atoms are close to 
Ref.~\cite{lambert_prb16}: $E_J/h=300$~GHz, $E_J/E_C=75$, $\alpha=0.7$ 
(relative area of the smaller junction). The model is extended to include the 
linear inductance, and the persistent-current artificial atoms are assumed to 
be square loops of size $3\text{ }\mu\text{m}\times 3\text{ }\mu\text{m}$ and 
use $0.2\text{ }\mu\text{m}\times 0.1\text{ }\mu\text{m}$ traces (width 
$\times$ height). The error in the side lengths of the loops is assumed to be 
$0.1\%$, and they are spaced $1\text{ }\mu\text{m}$ apart.
\label{fig_pcaa_parameters}}
\end{figure}

The distributions of the parameters in the above Hamiltonian were calculated 
using circuit quantization as described in App.~\ref{app_pcaa}. The input 
parameters to this calculation are written in the caption of 
Fig.~\ref{fig_pcaa_parameters} that also shows the distributions graphically. 
For the chosen bias flux $\Phi_\text{ext}$, the detunings $\Delta_j/(2\pi)$ 
have a standard deviation of $1.5$~GHz, and the average transition 
frequency is $\omega_{01,\text{avg}}/(2\pi)=8.0$~GHz. The average 
transition frequency $\omega_{12,\text{avg}}/(2\pi)=32.2$~GHz between 
the first and second excited states is only calculated to show that the 
anharmonicity is so large that the artificial atoms can be treated as 
two-level systems, and is not used in the 
Hamiltonian~\eqref{H_long_range_ising}.

Only the nearest-neighbor couplings $J_{\pm,\text{nn},\text{avg}}$ and 
$J_{z,\text{nn},\text{avg}}$ are shown Fig.~\ref{fig_pcaa_parameters}, but all 
the coupling strengths $J_{\pm,jk}$ and $J_{z,jk}$ in the 
Hamiltonian~\eqref{H_long_range_ising} are calculated based on the mutual 
inductance between the superconducting loops and then used in the simulations. 
The distribution of $J_{z,\text{nn}}/(2\pi)$ has average 
$J_{z,\text{nn},\text{avg}}/(2\pi)=39.8$~MHz and standard deviation 
$4.4$~MHz. The distribution of $J_{\pm,\text{nn}}/(2\pi)$ 
has average $J_{\pm,\text{nn},\text{avg}}/(2\pi)=6.1$~MHz and standard 
deviation $3.6$~MHz. We account for the small variations in the Rabi 
frequency for the different spins due to the inhomogeneity in the loop area by 
multiplying the ideal Rabi frequency $\Omega$ with time-independent factors 
$f_j$, which are are equal to the ratio of the actual area of the 
superconducting loop to the design value. They have the average value $1$ and 
standard deviation $0.0014$.

The idealized transverse-field Ising model of 
Refs.~\cite{lee_pra05,furman_prb06,close_prl11} is obtained when couplings in 
the Hamiltonian~\eqref{H_long_range_ising} are such that $J_{\pm,jk}=0$, and 
only the nearest-neighbor $J_{z,jk}$ are non-zero. Note that the scaling of 
the $\Delta_j$ and $J_{z,jk}$ terms in Hamiltonian~\eqref{H_long_range_ising} 
is different from Refs.~\cite{lee_pra05,furman_prb06,close_prl11}, and the 
choice of the scaling is different among those references. We also have the 
double counting of the terms $J_{z,jk} \hat{\sigma}_{z,j}\hat{\sigma}_{z,k}$ 
in the Hamiltonian~\eqref{H_long_range_ising}. This was chosen for consistency 
with the summation over the terms 
$J_{\pm,jk} \hat{\sigma}_{+,j}\hat{\sigma}_{-,k}$, where no double counting 
occurs, because the operators for $j$ and $k$ are different. Both summations 
arise from the summation over the off-diagonal elements of the inductance 
matrix which is symmetric. Non-zero $J_{\pm,\text{nn},\text{avg}}$ is always 
present in the persistent-current artificial atoms.

For the chosen parameters, the flux bias $\Phi_\text{ext}=0.5034\cdot \Phi_0$, 
where $\Phi_0$ is the magnetic flux quantum, minimizes the inhomogeneous 
broadening due to the variations in the Josephson junction 
area~\cite{lambert_prb16}. However, this is masked by the inhomogeneity due to 
variations in the loop area. While performing circuit quantization, small 
additional inhomogeneous broadening of $\Delta_j$ appears due to the 
long-range couplings, even in the absence of fabrication imperfections.

We consider two different initial states: the zero-excitation state 
\begin{gather}\label{psi_0_definition}
|\psi_0\rangle=\bigotimes_j |0_j\rangle,
\end{gather}
and the superposition of the single-excitation states
\begin{gather}\label{psi_1_definition}
|\psi_1\rangle = \frac{1}{\sqrt{N}}\sum_j\hat{\sigma}_{+,j}|\psi_0\rangle
\end{gather}
that can be a result of $N$ spins collectively absorbing a 
photon~\cite{lukin_rmp03,hammerer_rmp10}. Unless noted otherwise, the results 
below focus on such a delocalized excitation, but we also briefly consider the 
localized initial state 
$|\psi_1\rangle = \hat{\sigma}_{+,j=1}|\psi_0\rangle$, with $j=1$ being the 
corner spin~\cite{close_prl11}.
The measurement operator for the population (total number) of the excited spins is
$\hat{M}=\sum_{j}\hat{\sigma}_{+,j}\hat{\sigma}_{-,j}$.
Writing the spin amplification operation as a superoperator~$\mathcal{U}$, 
we thus maximize the expectation value
\begin{gather}\label{open_population_diff_def}
P_1-P_0=
\tr[\hat{M}\mathcal{U}(|\psi_1\rangle\langle\psi_1|)]
-\tr[\hat{M}\mathcal{U}(|\psi_0\rangle\langle\psi_0|)].
\end{gather}
This expectation value is the difference of excited-state populations and 
hence $-N\leq P_1-P_0 \leq N$.

For the optimal control, the main parametrization of the Rabi frequency 
$\Omega$ in the 
Hamiltonians~\eqref{H_long_range_ising_Rydberg}~and~\eqref{H_long_range_ising} 
that we use is based on the filtered piecewise-constant parametrization from 
Ref.~\cite{motzoi_pra11}. Compared to the unfiltered piecewise-constant 
parametrization where  penalty terms are added to make the pulses physically 
realizable, and the suitable constants multiplying the penalty terms are found 
by trial-and-error~\cite{heeres_ncomms17}, embedding the constraints in the 
pulse parametrization itself~\cite{motzoi_pra11} avoids the need for such 
tuning.

Gaussian filtering of the indicator functions
\begin{gather}
\mathbf{1}_{[a,b]}(t)=
\begin{cases}
1 ~&\text{ if } t\in[a,b],\\
0 ~&\text{ if } t\notin[a,b].
\end{cases}
\end{gather}
was experimentally verified to be a good model for the finite bandwidth of the 
arbitrary waveform generators~\cite{motzoi_pra11}. Our only change of this 
parametrization is to use $\text{tanh}$ function (that maps the real line onto 
the compact interval $(-1,1)$) on every amplitude to additionally constrain the 
extremal values. This results in
\begin{subequations}
\begin{align}
\begin{aligned}
\text{Re}[\Omega](t)=\frac{\Omega_\text{max}}{\sqrt{2\pi}\sigma}\int_{-\infty}^{\infty}
\exp\mathopen{}\left(-\frac{(t-t')^2}{2\sigma^2}\right)\\
\times\sum_{p=1}^{N_\text{c}} \text{tanh}(a_p)
\mathbf{1}_{[(t_\text{f}/N_\text{c})(p-1),(t_\text{f}/N_\text{c})p]}(t')
\dif t',
\end{aligned}
\label{Re_Omega_parametrization_constrained}
\\
\label{Im_Omega_parametrization_constrained}
\begin{aligned}
\text{Im}[\Omega](t)=\frac{\Omega_\text{max}}{\sqrt{2\pi}\sigma}\int_{-\infty}^{\infty}
\exp\mathopen{}\left(-\frac{(t-t')^2}{2\sigma^2}\right)\\
\times\sum_{p=1}^{N_\text{c}} \text{tanh}(b_p)
\mathbf{1}_{[(t_\text{f}/N_\text{c})(p-1),(t_\text{f}/N_\text{c})p]}(t')
\dif t',
\end{aligned}
\end{align}
\label{Re_Im_Omega_parametrization_constrained}
\end{subequations}
where $\Omega_{\text{max}}$ is the maximum absolute Rabi frequency.
The integrals in Eqs.~\eqref{Re_Im_Omega_parametrization_constrained} can be 
evaluated analytically in terms of the error functions~\cite{motzoi_pra11}. 
The parametrization~\eqref{Re_Im_Omega_parametrization_constrained} is for the 
persistent-current artificial atoms, where the detunings are constant in time, 
and the $I$ and $Q$ parts of the 
microwave pulse are shaped. For the Rydberg atoms, $\Omega$ 
is real and positive, and $\Delta$ is changed in time~\cite{omran_science19}. 
To make $\Omega$ positive, we use the logistic function 
$\sigma(x)=1/(1+e^{-x})$ (that maps the real line onto the interval $(0,1)$) 
instead of $\tanh$ in Eq.~\eqref{Re_Omega_parametrization_constrained}, 
and take Eq.~\eqref{Im_Omega_parametrization_constrained} to mean
the parametrization of $\Delta(t)$ instead of $\text{Im}[\Omega](t)$ 
(replacing $\Omega_\text{max}$ with $\Delta_\text{max}$).

For the persistent-current artificial atoms, we assume the sampling rate of 
$10$~GS/s~\cite{ding_prresearch24} that determines $N_\text{c}$ in 
Eqs.~\eqref{Re_Im_Omega_parametrization_constrained}, and the bandwidth 
$1/\sigma=4$~GHz. In the case of the Rydberg atoms, the output of the 
arbitrary waveform generator is fed into an acousto-optic 
modulator~\cite{omran_science19}, which determines the final bandwidth 
$1/\sigma$ of the pulses. We assume $1/\sigma=25$~MHz~\cite{aom_specs} 
in this case, and therefore the sampling rate is set to a lower value of $1$~GS/s. 
To ensure that $\Omega$ is close to zero for $t=0$ and $t=t_\text{f}$, we set 
a certain number of $a_p$ and $b_p$ to zero in the beginning and the end of 
the interval~\cite{motzoi_pra11}. For this, we use a heuristic rule that 
$a_p$ and $b_p$ are set to zero for $p-1$ below $4\sigma N_\text{c}/t_\text{f}+1$ and 
above $N_\text{c}-4\sigma N_\text{c}/t_\text{f}-1$. This 
results in $|\Omega|/(2\pi),|\Delta|/(2\pi) < 0.5$~kHz for $t=0$ and 
$t=t_\text{f}$ in the pulse shapes for all the plots below.

We could not reach the theoretical bound $P_1-P_0 = N$ with the 
parametrization~\eqref{Re_Im_Omega_parametrization_constrained}, even when the 
other imperfections (finite coherence time, inhomogeneity) were removed from 
the model. Therefore, to check that this theoretical bound is indeed 
attainable, we also use the Fourier series parametrization~\cite{doria_prl11}
\begin{subequations}
\begin{align}
&\text{Re}[\Omega_\text{Fourier}](t)=\sqrt{\frac{2}{t_\text{f}}}
\sum_{p=1}^{N_\text{c}} a_p \sin(\omega_p t),\displaybreak[0]\\
&\text{Im}[\Omega_\text{Fourier}](t)=\sqrt{\frac{2}{t_\text{f}}}
\sum_{p=1}^{N_\text{c}} b_p \sin(\omega_p t),
\end{align}
\label{Re_Im_Omega_parametrization}
\end{subequations}
where $N_\text{c}$ is the number of the Fourier components, and 
$\omega_p=p\pi/t_\text{f}$. The Rabi frequency is zero for the initial time 
$t=0$ and the final time $t=t_\text{f}$. We set $N_\text{c}=100$ so that it is 
large enough to parametrize pulses with bandwidth of several GHz for 
amplification times below $10$~ns.

The initial states are propagated in time with either the Schr\"odinger or 
master equation using the 4th order Runge-Kutta method. When evolving the 
master equation is too costly (many time steps or averaging over many 
inhomogeneity realizations), the ensembles with $N=16$ spins are simulated 
using the quantum trajectory method~\cite{daley_tandf14}. In all cases, the 
full Hilbert space with the basis size $2^N$ is used, necessitating 
significant computational power already for modestly large ensembles.

The detuning $\Delta_\text{avg}$ from the average spin frequency is optimized 
together with the pulse shape parameters $a_p$ and $b_p$. For the 
persistent-current artificial atoms, $\Delta_\text{avg}$ is one 
time-independent parameter, while for the Rydberg atoms, it is a 
time-dependent shape that is parametrized by the values $b_p$. Our optimal 
control approach is similar to Ref.~\cite{iakoupov_prresearch23}, where the 
gradient of the final population difference~\eqref{open_population_diff_def} 
with respect to the pulse parameters is found using the reverse-mode automatic 
differentiation [cf. App.~\ref{app_gradient}] and then used in a 
gradient-based optimization algorithm LBFGS~\cite{nocedal_moc80}.

In contrast to the GRAPE method~\cite{khaneja_jmr05} where finding the exact 
gradient is computationally expensive, and it is often approximated for 
efficiency~\cite{motzoi_pra11,chen_sciadv25}, the gradients of the Runge-Kutta 
methods can be calculated exactly with a small-constant overhead compared to 
the first-order (Euler method) 
approximation~\cite{gunther_qcs21,iakoupov_prresearch23}. We use the 4th order 
method similar to Ref.~\cite{iakoupov_prresearch23} instead of the 2nd order 
method of Ref.~\cite{gunther_qcs21}, because the higher order requires fewer 
time steps. In App.~\ref{app_reduced_storage}, we detail an almost matrix-free 
operator approach that is more efficient than the sparse-matrix operators in 
Refs.~\cite{iakoupov_prresearch23,gunther_qcs21} and can be used with any 
Runge-Kutta method. Matrix-free operators can also be used with the Krotov 
method~\cite{somloi_chemphys93}, but its quasi-Newton extension (similar to 
BFGS) is not straightforward~\cite{eitan_pra11}.

In our implementation, the total memory requirements for the optimal control 
are around $14.5$ ($18.5$) times the size of the density matrix (state 
vector). The difference between the multiples for the density matrix and the 
state vector, is that some auxiliary vectors have a fixed size equal to the 
state vector, and hence are negligible in size compared to the density matrix. 
Importantly, the memory requirements are independent of the number of the 
optimal control parameters and simulation time steps. The latter is achieved 
by propagating the final states backward in time during the gradient 
calculation~\cite{kosloff_chemphys89,somloi_chemphys93}. This makes it 
possible to simulate and optimize the control pulse shapes for ensembles of up 
to $N=16$ spins using the master equation.

The optimization method above is only guaranteed to converge to a local 
optimum, and we use the multistart approach to probabilistically explore more 
of the optimization landscape. For all the optimal control results, the pulse 
shapes are initialized randomly with each $a_p$ and $b_p$ sampled from the 
standard normal distribution (with mean $0$ and standard deviation $1$). Since 
different frequency scales are used for the two different kinds of spins that 
we consider, $\Delta_\text{avg}/(2\pi\cdot1\text{ MHz})$ is sampled from the 
standard normal distribution for the Rydberg atoms, and 
$\Delta_\text{avg}/(2\pi\cdot40\text{ MHz})$ is sampled for the 
persistent-current artificial atoms. We take $10$ such initial pulse 
realizations and pick the best optimized pulse shape as the result.

Rydberg atoms and persistent-current artificial atoms have significantly 
different couplings and coherence times. In both cases, the lifetime $T_1$ is 
relatively large, but the pure dephasing time $T_{\phi}$ is the limiting 
factor for the coherent operations. We set $T_1=175$ $\mu$s and $T_{\phi}=20$ 
$\mu$s for the Rydberg atoms~\cite{scholl_nature21}, and 
$T_{1}=1\text{ }\mu\text{s}$ and $T_{\phi}=15\text{ }\text{ns}$ for the 
persistent-current artificial atoms~\cite{chang_prapplied22}. The 
persistent-current artificial atoms have a smaller coherence time, but they 
also have around $40$ times larger interaction strengths and can be driven 
with stronger Rabi frequencies. This makes it possible to perform spin 
amplification much faster than with the Rydberg atoms.

In both cases, to perform the spin amplification approximately coherently, the 
amplification time $t_\text{f}$ is chosen smaller than $T_{\phi}$. We set 
$t_\text{f}=1\text{ }\mu\text{s}$ for the Rydberg atoms, and 
$t_\text{f}=6.5\text{ ns}$ for the persistent-current artificial atoms. For 
the latter, $t_\text{f}$ is close to the time scale set by the 
nearest-neighbor interaction $1/|J_{z,\text{nn}}| \approx 4\text{ ns}$. For 
the Rydberg atoms, $t_\text{f}$ is limited by the maximal Rabi frequencies 
possible in the experimental setups, and hence does not approach 
$1/|J_{z,\text{nn}}|$ as closely.

The previous spin amplification 
protocols~\cite{lee_pra05,furman_prb06,close_prl11,nill_prl24} used a small 
constant Rabi frequency $|\Omega|\ll |J_{z,\text{nn}}|$ that resulted in 
amplification times much longer than $1/|J_{z,\text{nn}}|$. This approach only 
works for long $T_{1}$ and $T_\phi$. Finite $T_\phi$ makes the spin 
amplification slower~\cite{close_prl11} and also introduces erroneous 
excitations in the initial zero-excitation case as explained in 
App.~\ref{app_cw_drive}. Finite $T_1$ limits the total achievable excitation 
number in the single-excitation case. While choosing a larger $|\Omega|$ could 
partially compensate for this, this also increases the zero-excitation error 
due to the finite $T_\phi$.

Since the largest contribution to the inhomogeneity in persistent-current 
artificial atoms stems from the geometric variations of the superconducting 
loops and hence is fixed after the fabrication, we optimize distinct pulse 
shapes for every inhomogeneity realization. In contrast, Rydberg atoms have a 
much smaller but dynamic inhomogeneity that stems from indeterminate 
positions of the atoms in the trap sites. Therefore, for the Rydberg atoms, 
the pulse shapes are optimized such that they work for any inhomogeneity 
realization with the average $P_1-P_0$ being the target to optimize.

\section{Results}\label{sec_results}

For a $2$D homogeneous ensemble with the nearest-neighbor couplings, a 
two-frequency continuous-wave drive is sufficient~\cite{close_prl11}. Since 
the considered systems are inhomogeneous with long-range couplings, we follow 
the recommendation of Ref.~\cite{furman_prb06} to increase the drive 
strengths, which may overcome frequency mismatch. This leads us to consider 
the baseline with the following time-dependent Rabi frequency 
\begin{gather}\label{two_frequency_Omega}
\Omega(t)=\Omega_{1}+\Omega_{2}
\exp(i(\Delta_{\text{avg},1}-\Delta_{\text{avg},2})t),
\end{gather}
where the detuning 
$\Delta_{\text{avg},1}=\omega_{\text{d},1}-\omega_{01,\text{avg}}$ of the 
first frequency component is included in 
$\Delta_j=\Delta_{\text{avg},1}+\omega_{{01},\text{avg}}-\omega_{01,j}$ in the 
Hamiltonian~\eqref{H_long_range_ising}.

\begin{figure}[t]
\begin{center}
\includegraphics{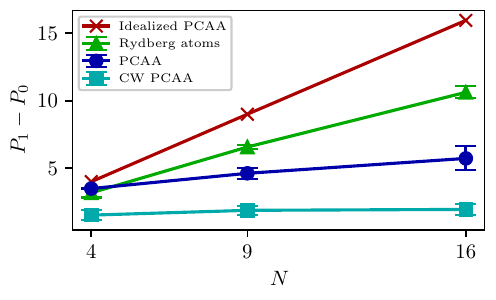}
\end{center}
\caption{Difference of the final populations $P_1-P_0$ for the spin 
amplification as a function of the number of spins~$N$. The results are for 
two types of spins: Rydberg atoms and persistent-current artificial atoms 
(PCAA). For the continuous-wave (CW) PCAA results (cyan squares), the 
two-frequency parametrization~\eqref{two_frequency_Omega} is used for the Rabi 
frequency with optimized~\cite{jones_jota93,nlopt} $\Omega_1$, $\Omega_2$, 
$\Delta_{\text{avg},1}$, and $\Delta_{\text{avg},2}$. The other curves show 
results for the drives found using optimal control. The idealized PCAA results 
(red crosses) are obtained using the Schr\"odinger equation with retained 
long-range couplings, but without decay, dephasing, and inhomogenity; with the 
pulse parametrization~\eqref{Re_Im_Omega_parametrization}. They lie very close 
to the line $P_1-P_0=N$. The results using optimal control for PCAA (blue 
circles) and Rydberg atoms (green triangles) are obtained with the master 
equation or the quantum trajectory method with the 
parametrization~\eqref{Re_Im_Omega_parametrization_constrained}. In all cases 
except the idealized PCAA, the inhomogeneity is modeled by averaging over a 
number of realizations, and the error bars show the standard deviation of this 
averaging. The number of inhomogeneity realizations is chosen to be $100$, 
except for CW PCAA with $N=16$ where only $15$ realizations are used, because 
each one of them needs a slow global optimization~\cite{jones_jota93,nlopt}.
\label{fig_ensemble_size}}
\end{figure}

Optimizing over $\Omega_{l}$ and $\Delta_{\text{avg},l}$ for $l=1,2$ in 
Eq.~\eqref{two_frequency_Omega} makes it a rudimentary form of optimal 
control. For every inhomogeneity realization of the persistent-current 
artificial atoms, we perform a global optimization with the algorithm 
DIRECT~\cite{jones_jota93,nlopt} inside the bounds 
$\Omega_{l}/(2\pi)\in[0.1,2.4]\cdot 40$~MHz, 
$\Delta_{\text{avg},l}/(2\pi)\in[-5,5]\cdot 40$~MHz, where $40$~MHz 
is approximately equal to the nearest-neighbor $|J_{z,jk}|/(2\pi)$ in 
Hamiltonian~\eqref{H_long_range_ising}. The average values of the population 
differences $P_1-P_0$ and error bars showing the standard deviations are shown 
as cyan squares in Fig.~\ref{fig_ensemble_size}. The population differences 
$P_1-P_0$ appear to reach a limit already for $N=16$ with the value 
$P_1-P_0=1.9\pm0.4$ (increasing less than $0.1$ from $N=9$). The optimal 
$\Omega/(2\pi)\approx |J_{z,\text{nn}}|/(2\pi)\approx 40$~MHz is outside of 
the parameter regime considered in 
Refs.~\cite{lee_pra05,furman_prb06,close_prl11}.

\begin{figure*}[t]
\begin{center}
\includegraphics{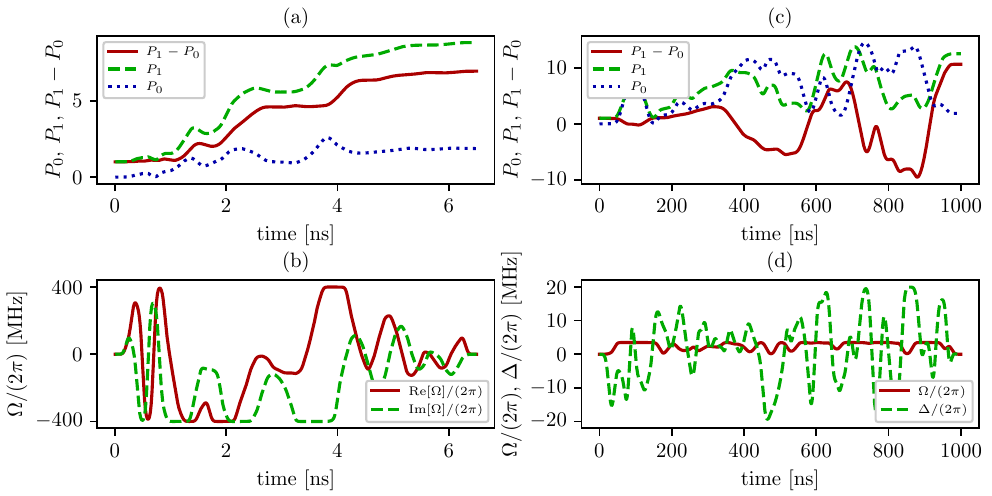}
\end{center}
\caption{(a) Populations as functions of time for one inhomogeneity 
realization of $N=16$ persistent-current artificial atoms as spins. (b) Pulse 
shapes that drive the dynamics in (a). The pulse shapes are parametrized such 
that the extremal values of $\text{Re}[\Omega]/(2\pi)$ and 
$\text{Im}[\Omega]/(2\pi)$ [cf. 
Eqs.~\eqref{Re_Im_Omega_parametrization_constrained}] are at $\pm 400$~MHz, 
and the filtering bandwidth is $1/\sigma=4$~GHz. (c) Average 
populations as functions of time for $100$ inhomogeneity realizations of 
$N=16$ Rydberg atoms as spins. (d) Pulse shapes that drive the dynamics in 
(c). The Rabi frequency $\Omega/(2\pi)$ is real and positive with the maximum 
value $3.5$~MHz. The detuning $\Delta/(2\pi)$ is constrained by the extremal 
values $\pm 20$~MHz.
\label{fig_pulse_shapes}}
\end{figure*}

Using optimal control with the 
parametrization~\eqref{Re_Im_Omega_parametrization_constrained} that has more 
frequency components and the gradient-based optimization, improves the 
performance of spin amplification over the baseline above. More importantly, 
$P_1-P_0$ increases with increasing $N$, even though the scaling is sublinear. 
The results are shown by blue circles in Fig.~\ref{fig_ensemble_size}. The 
values for $N=4$ and $N=9$ were optimized using the master equation directly. 
For $N=16$, a different approach was taken, and the value $P_1-P_0=5.7\pm 0.9$ 
is not fully optimized. The pulse shapes were found using the Schr\"odinger 
equation but with retained  parameter inhomogeneity, which is the most 
important imperfection. These pulse shapes were then evaluated using the 
master equation to obtain the blue circle for $N=16$ in 
Fig.~\ref{fig_ensemble_size}. Optimizing all of the $100$ inhomogeneity 
realizations with the master equation is too computationally expensive. On 
dual AMD Epyc 9554 CPUs ($128$ cores in total), one iteration takes around 
$90$~hours with $817$~simulation time steps. The number of time steps is 
double that of the homogeneous-ensemble results below to prevent numerical 
instability caused by the large inhomogeneity in the parameters. We have 
chosen $3$~realizations that were optimized for $19$~iterations each. This has 
improved $P_1-P_0$ for this subset of realizations from $5.9 \pm 0.7$ to 
$6.1 \pm 0.7$.

Besides the inhomogeneity in the system, the other important imperfection is 
the finite $T_\phi=15$ ns, which is not much larger than the chosen 
amplification time of $6.5$ ns. To quantify the influence of inhomogeneity in 
isolation from the other imperfections, pulse shapes were also optimized for a 
homogeneous ensemble with $N=16$. After $44$ iterations, each with $408$ time 
steps for speed, starting from a pulse shape optimized by the Schr\"odinger 
equation, $P_1-P_0=6.2$ was obtained, which is the same as the result for the 
inhomogeneous ensemble within the uncertainty. For the homogeneous ensemble, 
optimization with the master equation is more important. Using a pulse shape 
taken directly from the Schr\"odinger equation optimization and evaluating it 
with the master equation gives $P_1-P_0=4.3$, which is smaller than even the 
result obtained for the inhomogeneous ensemble with the same procedure. This 
suggests that optimizing the pulse shapes with the included inhomogeneity 
also makes them more robust against the decoherence, even when the latter is 
not explicitly included in the model.

Setting $T_1$ and $T_\phi$ to be infinite by using the Schr\"o\-din\-ger 
equation result for $N=16$, gives $P_1-P_0=8.0\pm 1.0$ with inhomogeneity or 
$P_1-P_0=14.4$ without inhomogeneity. The latter result is limited by the 
pulse shape constraints. Relaxing the constraints by using the pulse 
parametrization~\eqref{Re_Im_Omega_parametrization}, the theoretical upper 
bound $P_1-P_0=N$ is closely approached. This is shown by the red crosses in 
Fig.~\ref{fig_ensemble_size}. Even in this more idealized setting, optimal 
control with broadband pulses was found to be necessary, because of the 
retained long-range couplings and finite-size effects. Optimizing the 
continuous-wave parametrization~\eqref{two_frequency_Omega} and the 
amplification time (up to $800$ ns) for $N=16$, gives only $P_1-P_0=6.9$.

Using the pulse with a linear ramp of $\Delta$ inspired by the adiabatic state 
preparation~\cite{omran_science19} gives $P_1-P_0=10.0$ for the 
optimized~\cite{jones_jota93,nlopt} minimum and maximum 
$|\Delta|/(2\pi)\approx 9.7\cdot 40$~MHz and Rabi frequency 
$\Omega/(2\pi)\approx 1.2\cdot 40$~MHz. This is the only case where the 
$\Delta$-ramp pulse improves on the continuous-wave 
parametrization~\eqref{two_frequency_Omega}. With realistic dephasing and 
inhomogeneity, both for the persistent-current artificial atoms and the 
Rydberg atoms, the $\Delta$-ramp pulse gives worse results. A discussion of 
these pulses can also be found at the end of App.~\ref{app_cw_drive}.

The dynamics in time of the spin populations with the optimal control pulses 
is shown in Fig.~\ref{fig_pulse_shapes}(a), and the corresponding pulse shapes 
are shown in Fig.~\ref{fig_pulse_shapes}(b). The population difference 
$P_1-P_0$ does not necessarily increase monotonically in contrast to the weak 
continuous-wave spin amplification~\cite{lee_pra05,furman_prb06,close_prl11}. 

The above results for the persistent-current artificial atoms were using the delocalized 
single excitation given by Eq.~\eqref{psi_1_definition}. We have also 
checked the performance with the localized excitation $|\psi_1\rangle = \hat{\sigma}_{+,j=1}|\psi_0\rangle$, 
where $j=1$ is the corner spin~\cite{close_prl11}. For $N=16$ 
and using the same optimization procedure as for the 
state~\eqref{psi_1_definition}, $P_1-P_0=7.0\pm 1.1$ is obtained. This is
larger than $P_1-P_0=5.7\pm 0.9$ obtained for the 
state~\eqref{psi_1_definition}, and suggests that amplifying delocalized 
states is more challenging.

The results for the Rydberg atoms are shown by green triangles in 
Fig.~\ref{fig_ensemble_size}. The optimization for $N=16$ is done in the same 
way as for the persistent-current artificial atoms. First, the pulse shapes 
are found using the Schr\"odinger equation that includes the inhomogeneity 
[cf. Fig.~\ref{fig_pulse_shapes}(d)], and then they are evaluated using the 
quantum trajectory method with $T_1=175$ $\mu$s and $T_{\phi}=20$ $\mu$s [cf. 
Fig.~\ref{fig_pulse_shapes}(c)], resulting in $P_1-P_0=10.6\pm0.4$ for $N=16$. 
This value for $P_1-P_0$ is more limited by the pulse shape constraints 
compared to the persistent-current artificial atoms.

The optimal control result for the Rydberg atoms can be compared to the 
optimized~\cite{jones_jota93,nlopt} continuous-wave 
parametrization~\eqref{two_frequency_Omega} (within the bounds 
$\Omega_{l}/(2\pi)\in[0.1,2.4]\cdot 1$~MHz, 
$\Delta_{\text{avg},l}/(2\pi)\in[-5,5]\cdot 1$~MHz to account for the 
changed frequency scales) that gives $P_1-P_0=6.5\pm0.7$ for the amplification 
time of $1.1$~$\mu$s (up to $5$~$\mu$s was considered). Using a pulse with a 
linear ramp of $\Delta$ inspired by the adiabatic state 
preparation~\cite{omran_science19} gives a worse $P_1-P_0=5.6\pm0.1$ for the 
optimized~\cite{jones_jota93,nlopt} minimum and maximum 
$|\Delta|/(2\pi)\approx 7.4$~MHz, Rabi frequency 
$\Omega/(2\pi)\approx 0.5$~MHz, and ramping time of $20$~$\mu$s, where 
the maximum $P_1-P_0$ is achieved for the evolution time $6.7$~$\mu$s.

\section{Discussion and conclusion}\label{sec_discussion_conclusion}

Spin amplification is both useful in itself for quantum sensing or 
measurement, and as a step on the way to the universal quantum computation 
with a global driving field. For the quantum measurement of a single spin 
excitation with a global drive, spin amplification could be used before the 
global excitation is measured. In the above, we have considered 
persistent-current artificial atoms operated away from the degeneracy point as 
spins. In this case, the two spin states are associated with different 
magnetic fields that can be distinguished by a magnetometer. A superconducting 
quantum interference device (SQUID) fabricated on the same chip such that it 
encloses the entire ensemble, could be used as such a 
magnetometer~\cite{hatridge_prb11}. For the Rydberg atoms, one of the 
detection mechanisms relies on the selective trapping of the two spin states: 
atoms in the ground states are trapped and imaged, while the atoms in the 
excited (Rydberg) states are repelled from the trapping 
sites~\cite{ebadi_nature21,scholl_nature21}.

The single excitation that is amplified could be a result of an absorption of 
a single photon. In this case, spin amplification can be viewed as having the 
same function as the internal dynamics of the single-photon avalanche 
detectors~\cite{ceccarelli_aqt21,nill_prl24}. This can enable photodetectors 
in the frequency ranges where the technology is less developed. 
Superconducting artificial atoms as spins could detect the microwave photons, 
and Rydberg atoms could detect the terahertz photons~\cite{nill_prl24}.

Our results show that optimal control can significantly reduce the time needed 
for the amplification. For the Rydberg atoms, Ref.~\cite{nill_prl24} found 
that amplification in a 1D array of $11$ spins could be done in $25$ $\mu$s 
using weak continuous-wave drives, while we show that amplification in a 2D 
array of $16$ spins could be done in $1$ $\mu$s using optimal control, either 
a simpler form that uses two optimized frequency components, or with a 
broadband pulse shape that achieves better amplification. Since the dephasing 
time $T_{\phi}$ of the Rydberg levels is around $20$~$\mu$s~\cite{scholl_nature21}, 
much faster amplification time using optimal control is the key improvement 
that makes spin amplification implementable using these spins. For the 
persistent-current artificial atoms, our calculations show that amplification 
time of $6.5$ ns is possible using optimal control, which also makes it 
shorter than $T_{\phi}=15$ ns, providing another example that faster 
operations using optimal control are essential for spin amplification in 
realistic spins.

Universal quantum computation with globally driven spin ensembles is also 
possible~\cite{benjamin_pra00,benjamin_prl01,fitzsimons_prl06,cesa_prl23,patomaki_npjqi24}.
The techniques that we have developed for the analysis of spin amplification, 
could be used to better understand which other operations can be performed 
with such setups even in the presence of imperfections.

It is possible to have ensembles with much larger numbers of spins in the 
experimental setups of 
Refs.~\cite{ebadi_nature21,scholl_nature21,kakuyanagi_prl16} than we have 
considered. To find pulse shapes for such ensembles theoretically, the direct 
simulation of the exponentially-large Hilbert space is impossible. Alternative 
approaches, such as the matrix-product 
states~\cite{ebadi_nature21,scholl_nature21,muleady_prl23} or semiclassical 
methods~\cite{muleady_prl23}, will be required. Experimentally, the pulses 
could also be found by optimizing the setup 
directly~\cite{werninghaus_npjqi21}. The latter approach will be required to 
optimize the pulses for every static inhomogeneity realization as we have 
assumed for the persistent-current artificial atoms. For the Rydberg 
atoms, where the inhomogeneity is significantly smaller, the pulse shapes 
found theoretically can be applied to any inhomogeneity realization, and thus 
it should be possible to use them as a starting point for implementing spin 
amplification in the experimental setups.

To conclude, our calculations show that spin amplification is possible in 
noisy spin ensembles consisting of superconducting persistent-current 
artificial atoms or Rydberg atoms. This is done by employing optimal control 
techniques to find classical drive pulse shapes. Thereby the time of the spin 
amplification is significantly shortened compared to the previous protocols 
that use weak continuous-wave drives.

\begin{acknowledgments}
The data that support the findings of this study were generated by numerical 
simulations. The data and the source code used to generate the data are 
publicly available~\cite{zenodo}. We thank Kosuke Kakuyanagi for useful 
discussions. We are grateful for the help and support provided by the 
Scientific Computing and Data Analysis section of Core Facilities at OIST.
\end{acknowledgments}

\appendix
\section{Persistent-current artificial atoms}\label{app_pcaa}
In this appendix, we show how the spin Hamiltonian~\eqref{H_long_range_ising} 
can be derived from the circuit quantization of the persistent-current 
artificial atoms~\cite{orlando_prb99}. We need to include the self-inductance 
terms in the artificial atom 
Hamiltonian~\cite{maassen_prb05,robertson_prb06,yamamoto_njp14,yan_ncomms16}, 
because they significantly modify the transition frequencies and are needed to 
introduce the couplings that are caused by the mutual inductance. The 
inductance matrix $\mathbf{L}$ that combines both the self-inductances 
(diagonal elements) and the mutual inductances (off-diagonal elements) is 
calculated using a version of FASTHENRY~\cite{kamon_ieee94} that includes the 
kinetic inductance due to superconductivity~\cite{xictools}. 

For the calculation of the kinetic inductance, we set the London penetration 
depth $\lambda=0.2\text{ }\mu\text{m}$ as an order-of-magnitude estimate, in 
line with the measurements of the aluminum thin films~\cite{steinberg_prb08}. 
The loops of the persistent-current artificial atoms are set to be squares 
with the design value for the side length equal to $3$ $\mu$m, but due to 
finite accuracy of the electron-beam lithography, these lengths have an error, 
which we set to be $3$ nm~\cite{duan_jvstb10}. The centers of the artificial 
atoms are assumed to be fixed, and it is thus only their loop areas that are 
changing between the inhomogeneity realizations. For every such realization, a 
different input file to FASTHENRY is generated with the side lengths sampled 
from a normal distribution with the average value $3$ $\mu$m and standard 
deviation $3$ nm. The other parameters are given in the caption of 
Fig.~\ref{fig_pcaa_parameters}. We get the self-inductance $L=37.45\pm 0.03$~pH 
and the mutual inductance of the nearest neighbors $M=-0.2187\pm 0.0006$~pH. 
The factors $f_j$ in the Hamiltonian~\eqref{H_long_range_ising} are equal to 
the ratio of the actual area of the superconducting loop to the design value, 
and $\Omega$ is the time-dependent Rabi frequency that assumes the coupling to 
a loop with the design value of the area.

\begin{figure}[t]
\begin{center}
\includegraphics{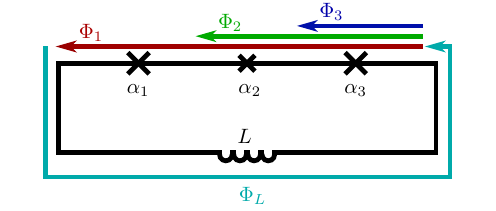}
\end{center}
\caption{Flux variable definitions. The integration contours that correspond 
to the flux variables are displaced out of the circuit for clarity, since some 
of them overlap.
\label{fig_flux_definitions}}
\end{figure}

Before the Hamiltonian can be found, a classical Lagrangian $\mathcal{L}_j$ 
for the persistent-current artificial atom $j$ is determined from the flux 
variables shown in Fig.~\ref{fig_flux_definitions}. The index $j$ is omitted 
on the flux variables for brevity. The fluxes are time integrals of voltages 
$\Phi_C(t)=\int_{-\infty}^t V_C(t')\dif t'$, and the voltages are electric 
field line integrals 
$V_C = \int_C \vec{E}\cdot\dif \vec{l}$~\cite{devoret_1997,vool_cta17} 
along the different contours $C$ that are segments of the loop, as shown in 
Fig.~\ref{fig_flux_definitions}. There is gauge 
freedom~\cite{devoret_1997,vool_cta17} in the sense that different choices of 
the contours $C$ are 
possible~\cite{robertson_prb06,yamamoto_njp14,yan_ncomms16}, but they all lead 
to the same quantum energy levels. The total inductive energy in the ensemble 
is $\vec{\Phi}_{L}^\text{T}\mathbf{L}^{-1}\vec{\Phi}_{L}$~\cite{vool_cta17}, 
where $\vec{\Phi}_{L}$ is a vector with $N$ elements for the fluxes that 
correspond to the linear inductances in each artificial atom, and 
$\mathbf{L}^{-1}$ is the inverse of the inductance matrix. We include the 
diagonal terms of this expression that are proportional to 
$(\mathbf{L}^{-1})_{jj}$ in the potential energies of each artificial atom. 
The off-diagonal terms are subsequently projected onto the eigenstates to give 
the couplings between the artificial atoms.

We have $\mathcal{L}_j=T_j-U_j$, where the kinetic energy is
\begin{gather}
T_j=\frac{\alpha_1 C_\text{J}}{2}\left(\dot{\Phi}_1-\dot{\Phi}_2\right)^2
+\frac{\alpha_2 C_\text{J}}{2}\left(\dot{\Phi}_2-\dot{\Phi}_3\right)^2
+\frac{\alpha_3 C_\text{J}}{2}\dot{\Phi}_3^2,
\end{gather}
the potential energy is
\begin{gather}
\begin{aligned}
&U_j=(\mathbf{L}^{-1})_{jj}\frac{\Phi_L^2}{2}
-\alpha_1 E_\text{J}\cos\left(\frac{2\pi}{\Phi_0}(\Phi_1-\Phi_2)\right)\\
&-\alpha_2 E_\text{J}\cos\left(\frac{2\pi}{\Phi_0}(\Phi_2-\Phi_3)\right)
-\alpha_3 E_\text{J}\cos\left(\frac{2\pi}{\Phi_0}\Phi_3\right),
\end{aligned}
\end{gather}
and ${\Phi_0=h/(2e)\approx 2.068\times 10^{-15}}$~Wb is the magnetic flux 
quantum. The Josephson junction energies $E_\text{J}$ and capacitances $C_\text{J}$ are 
proportional to the Josephson junction area and are scaled by the
factors~$\alpha_l$. In the ideal case, $\alpha_1=\alpha_3=1$ and $\alpha_2=\alpha=0.7$, 
but due to fabrication imperfections the actual values deviate from the ideal 
ones. For the calculations with inhomogeneity, the values of $\alpha_l$ are 
sampled from a normal distribution with the average value $\mu$ equal to the 
ideal one, and the standard deviation $\sigma=0.02$.

The flux variables are not independent, as there is a constraint on the total 
flux through the loop given by
\begin{gather}
f_j\Phi_\text{ext}=\int_{-\infty}^t\oint \vec{E}(t')\cdot\dif \vec{l}\dif t'
=\Phi_1+\Phi_L,
\end{gather}
where $\Phi_\text{ext}$ is external flux defined to be the value for the 
design loop area, and the scale factors $f_j$ (the same as in 
Hamiltonian~\eqref{H_long_range_ising}) are used to model the loop area 
inhomogeneity. This is the dominant contribution to the standard deviation of 
$\Delta_j$ in Hamiltonian~\eqref{H_long_range_ising}, and it is an order of 
magnitude larger than the contribution due to the standard deviation of the 
factors $\alpha_l$.

The flux variable $\Phi_1$ can be eliminated using the constraint above.
Assuming external flux that is constant in time, $\dot{\Phi}_\text{ext}=0$, 
the kinetic energy is
\begin{gather}
T_j=\frac{\alpha_1 C_\text{J}}{2}\left(\dot{\Phi}_L+\dot{\Phi}_2\right)^2
+\frac{\alpha_2 C_\text{J}}{2}\left(\dot{\Phi}_2-\dot{\Phi}_3\right)^2
+\frac{\alpha_3 C_\text{J}}{2}\dot{\Phi}_3^2,
\end{gather}
and the potential energy is
\begin{gather}
\begin{aligned}
&U_j=(\mathbf{L}^{-1})_{jj}\frac{\Phi_L^2}{2}
-\alpha_1 E_\text{J}
\cos\left(\frac{2\pi}{\Phi_0}(f_j\Phi_\text{ext}-\Phi_L-\Phi_2)\right)\\
&-\alpha_2 E_\text{J}\cos\left(\frac{2\pi}{\Phi_0}(\Phi_2-\Phi_3)\right)
-\alpha_3 E_\text{J}\cos\left(\frac{2\pi}{\Phi_0}\Phi_3\right).
\end{aligned}
\end{gather}
The kinetic energy can be written in terms of the capacitance matrix
\begin{gather}
\mathbf{C}=C_\text{J}
\begin{pmatrix}
\alpha_1+\alpha_2 & -\alpha_2 & \alpha_1 \\
-\alpha_2 & \alpha_2+\alpha_3 &  0\\
\alpha_1 & 0 & \alpha_1
\end{pmatrix}
\end{gather}
as $T_j=(1/2)\dot{\vec{\Phi}}^\text{T} \mathbf{C} \dot{\vec{\Phi}}$, where 
$\vec{\Phi}^\text{T}=(\Phi_2, \Phi_3, \Phi_L)$  is 
the transpose of $\vec{\Phi}$.

The next step is to perform the Legendre transformation and thereby find the 
classical Hamiltonian. For the Legendre transformation, the vector of the 
canonical momenta is determined as 
\begin{gather}
\vec{Q}=\pd{\mathcal{L}_j}{\dot{\vec{\Phi}}}=\mathbf{C}\dot{\vec{\Phi}}.
\end{gather}
The classical Hamiltonian is then
\begin{gather}
H_j=\vec{Q}^\text{T}\dot{\vec{\Phi}}-\mathcal{L}_j=2T_j-\mathcal{L}_j=T_j+U_j,
\end{gather}
where the kinetic energy written in terms of the canonical momenta is 
$T_j=(1/2)\vec{Q}^\text{T}\mathbf{C}^{-1}\vec{Q}$.

Performing the canonical quantization, the Poisson bracket 
$\{\Phi_l,Q_m\}=\delta_{l,m}$ is replaced with the commutator 
$[\hat{\Phi}_l,\hat{Q}_m]=i\hbar\delta_{l,m}$. Defining the dimensionless 
external flux $f=\Phi_\text{ext}/\Phi_0$ and operators
\begin{align}
\hat{n}_l=\frac{1}{2e}\hat{Q}_l,&&
\hat{\varphi}_l=\frac{2\pi}{\Phi_0}\hat{\Phi}_l,
\end{align}
with the commutator $[\hat{\varphi}_l,\hat{n}_{m}]=i\delta_{l,m}$,
the kinetic energy can be written 
$\hat{T}_j=4E_C\vec{n}^\text{T} \mathbf{C}^{-1}\vec{n}$ with $E_C=e^2/(2C_J)$. The 
Hamiltonian is
\begin{gather}
\begin{aligned}\label{H_pcaa_L}
\hat{H}_j=&\;\hat{T}_j+\frac{1}{2}E_L\hat{\varphi}_L^2
-\alpha_1 E_\text{J}\cos\left(2\pi f_jf-\hat{\varphi}_L-\hat{\varphi}_2\right)\\
&-\alpha_2 E_\text{J}\cos\left(\hat{\varphi}_2-\hat{\varphi}_3\right)
-\alpha_3 E_\text{J}\cos\left(\hat{\varphi}_3\right),
\end{aligned}
\end{gather}
where $E_L=(\mathbf{L}^{-1})_{jj}(\Phi_0/(2\pi))^2$.

\begin{figure}[t]
\begin{center}
\includegraphics{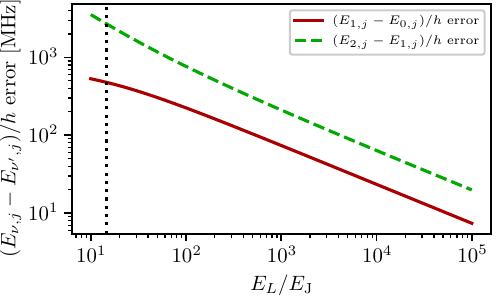}
\end{center}
\caption{The error in calculating the transition frequencies 
$(E_{1,j}-E_{0,j})/h$ and $(E_{2,j}-E_{1,j})/h$ for a persistent-current 
artificial atom $j$ using the zero-self-inductance 
Hamiltonian~\eqref{H_pcaa_no_L}, compared to the Hamiltonian~\eqref{H_pcaa_L} 
that accounts for the finite self-inductance. The error is plotted as a 
function of the ratio of the inductive energy $E_L$ and Josephson energy $E_\text{J}$. 
For the Hamiltonian~\eqref{H_pcaa_L}, the ideal values 
$\alpha_1=\alpha_3=1$, $\alpha_2=\alpha$ and $f_j=1$ are used. The bias flux 
is $f=\Phi_\text{ext}/\Phi_0=0.5034$ in both cases. Other parameters are 
stated in the caption of Fig.~\ref{fig_pcaa_parameters}, but the Josephson 
junction area error is not included. Assuming that the artificial atom is 
isolated, i.e., the inductance matrix $\mathbf{L}$ has only one element 
$L=37.4$~pH, we have $E_L/E_\text{J}=14.5$, which is shown by the black dotted 
vertical line. \label{fig_pcaa_E_L}}
\end{figure}

In the limit $L\rightarrow 0$ ($E_L \rightarrow \infty$), the constraint 
$\hat{\varphi}_L=0$ is imposed. Replacing 
$\hat{\varphi}_2\rightarrow\hat{\varphi}_2-\hat{\varphi}_1$ and 
$\hat{\varphi}_3\rightarrow-\hat{\varphi}_1$, and using the ideal 
$\alpha_1=\alpha_3=1$, $\alpha_2=\alpha$ and $f_j=1$, the Hamiltonian reduces 
to the one considered in Ref.~\cite{orlando_prb99}:
\begin{gather}\label{H_pcaa_no_L}
\begin{aligned}
\hat{H}_{j,0}=&\;T_{j,0}-\big(E_J\cos(\hat{\varphi}_1)+E_J\cos(\hat{\varphi}_2)\\
&+\alpha E_J\cos(2\pi f + \hat{\varphi}_1 - \hat{\varphi}_2)\big).
\end{aligned}
\end{gather}
The end result of the Hamiltonian diagonalization can be written as 
$\hat{H}_j=\sum_{\nu=0}^\infty E_{\nu,j} |\nu_j\rangle\langle \nu_j|$ in terms 
of the eigenenergies~$E_{\nu,j}$ and the eigenvectors~$|\nu_j\rangle$. For a 
finite $L$, there is a significant difference in the spectrum between the 
Hamiltonians $\hat{H}_{j,0}$ and $\hat{H}_j$, as shown in 
Fig.~\ref{fig_pcaa_E_L}. Our parameters approximately correspond to 
$E_L/E_J=14.5$ (ignoring the mutual inductance), and for this value the 
Hamiltonian~\eqref{H_pcaa_L} gives $(E_{1,j}-E_{0,j})/h=7.9$~GHz, while the 
Hamiltonian~\eqref{H_pcaa_no_L} gives $(E_{1,j}-E_{0,j})/h=8.4$~GHz.

For the case of finite but small $L$ that we consider, the term 
$(1/2)E_L\hat{\varphi}_L^2$ is dominant in the Hamiltonian~\eqref{H_pcaa_L}. 
To reduce the total number of the basis states required to diagonalize the 
entire Hamiltonian~\eqref{H_pcaa_L}, a harmonic oscillator basis is chosen 
where the quadratic part
\begin{gather}\label{H_pcaa_L_quadratic_part}
\hat{H}_L = A\hat{n}_L^2 + B \hat{\varphi}_L^2
\end{gather}
is already diagonal. In Eq.~\eqref{H_pcaa_L_quadratic_part}, $A$ is 
proportional to the element of the inverse capacitance matrix $\mathbf{C}^{-1}$ 
corresponding to $\hat{\varphi}_L$, and $B=(1/2)E_L$. The operators are 
written 
$\hat{\varphi}_L = \frac{1}{C_L}\frac{1}{\sqrt{2}}(\hat{a}+\hat{a}^\dagger)$, 
and $\hat{n}_L = -iC_L\frac{1}{\sqrt{2}}(\hat{a}-\hat{a}^\dagger)$, where the 
creation operator $\hat{a}^\dagger$ and the annihilation operator $\hat{a}$ 
are defined by their actions on the Fock states~$|n_L\rangle$. We have 
$\hat{a}^\dagger|n_L\rangle = \sqrt{n_L + 1} | n_L + 1\rangle$ and 
$\hat{a}|n_L\rangle = \sqrt{n_L} |n_L - 1\rangle$. The commutator 
$[\hat{\varphi}_L,\hat{n}_L]=i$ is satisfied for any choice of the constant 
$C_L$, but to diagonalize Eq.~\eqref{H_pcaa_L_quadratic_part}, a particular 
$C_L=(B/A)^{1/4}$ is chosen, resulting in 
$\hat{H}_L = 2\sqrt{AB}\del{\hat{a}^\dagger \hat{a} + \frac{1}{2}}$.

For the representation of the Josephson-junction operators in (phases 
$\hat{\varphi}_2$ and $\hat{\varphi}_3$, and the corresponding charge numbers 
$\hat{n}_2$ and $\hat{n}_3$ in Eq.~\eqref{H_pcaa_L}), the charge basis is 
used. In this basis, the eigenstates are $|n_l\rangle$ where $l=2,3$; and the 
operators can be written 
$\hat{n}_l = \sum_{n_l=-\infty}^\infty n_l |n_l\rangle\langle n_l|$, and 
$e^{-i\hat{\varphi}_l}=\sum_{n_l=-\infty}^\infty |n_l\rangle\langle n_l-1|$. 
In the harmonic oscillator basis, $e^{-i\hat{\varphi}_L}$ is computed using 
matrix exponentiation. The Hilbert spaces of the Josephson junctions are 
truncated to $21$ charge states each, and the harmonic oscillator basis to $8$ 
Fock states.

For the numerical diagonalization of Eq.~\eqref{H_pcaa_L}, the matrix 
exponentials are decomposed using tensor products. E.g., the expression 
$e^{i\hat{\varphi}_2 - i\hat{\varphi}_3}$ in Eq.~\eqref{H_pcaa_L} is actually
\begin{gather}\label{exp_phi_term_decomposition}
\begin{aligned}
&\exp\delnospace{i \hat{\varphi}_2\otimes\hat{I}_3\otimes \hat{I}_L 
- i\hat{I}_2\otimes \hat{\varphi}_3 \otimes \hat{I}_L}\\
&=\exp\delnospace{i\hat{\varphi}_2}\otimes \exp\delnospace{-i\hat{\varphi}_3}
\otimes \hat{I}_L
\end{aligned}
\end{gather}
where 
$\hat{I}_2$, $\hat{I}_3$ and $\hat{I}_L$ are the identity operators on the 
respective Hilbert spaces. In Eq.~\eqref{exp_phi_term_decomposition}, we have 
used the fact that $e^{\hat{A}+\hat{B}}=e^{\hat{A}}e^{\hat{B}}$ for 
commuting matrices $\hat{A}$ and $\hat{B}$, and that
\begin{gather}
\exp(\hat{A}\otimes \hat{I})
= \exp(\hat{A})\otimes \hat{I}.
\end{gather}
Similarly, the other operators in Eq.~\eqref{H_pcaa_L} are rewritten with the
identity operators appearing explicitly. E.g., $\hat{Q}_2^2$ becomes
$\hat{Q}_2^2\otimes \hat{I}_3 \otimes \hat{I}_L$. It is the latter types of 
expressions that are used to assemble the matrix for the numerical diagonalization.

Coupling between the artificial atoms with Hamiltonians $\hat{H}_j$ 
is due to the off-diagonal elements of the inverse inductance 
matrix $\mathbf{L}^{-1}$. The Hamiltonian for the entire ensemble can be 
written
\begin{align}\label{H_interaction}
\hat{H}_\text{ensemble}
=\sum_j \hat{H}_j
+\sum_{\substack{j,k\\j\neq k}}(\mathbf{L}^{-1})_{jk}
\hat{\Phi}_{L,j}\hat{\Phi}_{L,k}.
\end{align}
Making the two-level approximation, the operators $\hat{\Phi}_{L,j}$ are 
projected onto the subspace spanned by the eigenvectors $|0_j\rangle$ and 
$|1_j\rangle$ of $\hat{H}_j$ and hence can be written as the Hermitian 
$2\times 2$ matrices
\begin{gather}
\hat{\Phi}_{L,j}=
\begin{pmatrix}
\langle 0_j|\hat{\Phi}_{L,j}| 0_j\rangle &
\langle 0_j|\hat{\Phi}_{L,j}| 1_j\rangle \\
\langle 1_j|\hat{\Phi}_{L,j}| 0_j\rangle &
\langle 1_j|\hat{\Phi}_{L,j}| 1_j\rangle
\end{pmatrix}.
\end{gather}
Such matrices can be decomposed into linear combinations of Pauli matrices and 
the identity matrix $\hat{I}_j$. Thus,
\begin{gather}\label{Phi_Pauli_decomposition}
\hat{\Phi}_{L,j}
=P_{\hat{I}_j}
+P_{\hat{\sigma}_{x,j}}\hat{\sigma}_{x,j}
+P_{\hat{\sigma}_{y,j}}\hat{\sigma}_{y,j}
+P_{\hat{\sigma}_{z,j}}\hat{\sigma}_{z,j},
\end{gather}
where $P_{\hat{O}_j}$ for 
$\hat{O}_j\in\{\hat{I}_j,\hat{\sigma}_{x,j},\hat{\sigma}_{y,j},\hat{\sigma}_{z,j}\}$
are real coefficients found via the Hilbert-Schmidt inner product
\begin{gather}
P_{\hat{O}_j}=\frac{1}{2}\text{tr}\left[\hat{\Phi}_{L,j}\hat{O}_j\right].
\end{gather}
The eigenvectors are defined up to a phase, and the relative phases of the 
lowest-energy states $|0_j\rangle$ and $|1_j\rangle$ are chosen such that 
$\langle 0_j|\hat{\Phi}_{L,j}| 1_j\rangle$ is real. This results in 
$P_{\hat{\sigma}_{y,j}}=0$, but we still keep it in the expressions for 
generality.

Inserting Eq.~\eqref{Phi_Pauli_decomposition} into Eq.~\eqref{H_interaction},
applying the rotating wave approximation, and removing constants, we get
\begin{gather}
\begin{aligned}
\hat{H}_\text{ensemble}
&=\hbar\sum_{j} \frac{\omega_{j}}{2} \hat{\sigma}_{z,j}
+\hbar\sum_{\substack{j,k\\j\neq k}} J_{z,jk} \hat{\sigma}_{z,j}\hat{\sigma}_{z,k}\\
&+\hbar\sum_{\substack{j,k\\j\neq k}} J_{\pm,jk} \hat{\sigma}_{+,j}\hat{\sigma}_{-,k}
\end{aligned}
\end{gather}
with 
\begin{align}
&\omega_j=\frac{E_{1,j}-E_{0,j}}{\hbar}
+4\sum_{\substack{k\\k\neq j}}\frac{(\mathbf{L}^{-1})_{jk}}{\hbar}
P_{\hat{\sigma}_{z,j}}P_{\hat{I}_k},\label{omega_j_definition}\\
&J_{z,jk}=\frac{(\mathbf{L}^{-1})_{jk}}{\hbar}P_{\hat{\sigma}_{z,j}}
P_{\hat{\sigma}_{z,k}},\\
&J_{\pm,jk}=2\frac{(\mathbf{L}^{-1})_{jk}}{\hbar}
(P_{\hat{\sigma}_{x,j}}-iP_{\hat{\sigma}_{y,j}})
(P_{\hat{\sigma}_{x,k}}+iP_{\hat{\sigma}_{y,k}}).
\end{align}
The coupling $J_{\pm,jk}$ is real because of $P_{\hat{\sigma}_{y,j}}=0$.
Adding the driving terms and going into the rotating frame with respect to 
$\hat{H}_0=(\hbar\omega_\text{d}/2)\sum_j \hat{\sigma}_{z,j}$ results in the 
Hamiltonian~\eqref{H_long_range_ising} of the main text. Note that 
Eq.~\eqref{omega_j_definition} shows that even without fabrication 
imperfections, there is a small inhomogeneous broadening in the renormalized 
spin frequencies due to the fact that spins experience different sum of mutual 
inductances between them and the neighbors.

\section{Reduced-storage operators}\label{app_reduced_storage}
In this appendix, we show how the action of the quantum mechanical operators 
and superoperators can be calculated with reduced storage requirements. The 
state vectors $\psi$ are propagated in time using the Schr\"odinger equation
\begin{gather}\label{schroedinger_equation}
\dot{\psi}=-\frac{i}{\hbar}\hat{H}\psi,
\end{gather}
where the Hamiltonian $\hat{H}$ is given by either 
Eq.~\eqref{H_long_range_ising_Rydberg} or Eq.~\eqref{H_long_range_ising}, 
and the ket notation on the state vectors is omitted.
The density matrices $\rho$ are propagated in time using 
the Markovian zero-temperature master equation
\begin{gather}\label{master_equation}
\dot{\rho}=-\frac{i}{\hbar}[\hat{H},\rho]
+\gamma\sum_j\mathcal{D}[\hat{\sigma}_{-,j}]\rho
+\gamma_\text{d}\sum_j\mathcal{D}[\hat{\sigma}_{z,j}]\rho,
\end{gather}
where
$\mathcal{D}[\hat{o}]\rho=\frac{1}{2}(2\hat{o}\rho\hat{o}^\dagger
-\rho\hat{o}^\dagger\hat{o}-\hat{o}^\dagger\hat{o}\rho)$.

For an ensemble of two-level systems, there is a direct 
mapping between binary numbers and basis states. The 
zero-excitation state $\psi_0$ [Eq.~\eqref{psi_0_definition}] is 
mapped to the binary number $0$. A basis state 
$\hat{\sigma}_{+,j}\hat{\sigma}_{+,k}\psi_0$ that has spins $j$ and $k$ 
up and all other spins down, is mapped to the binary number where bits $j$ and 
$k$ are set (equal to $1$), and all the other bits are reset (equal to $0$). 
Interpreted as an integer, this number has value $2^j+2^{k}$ (if $j$ and $k$ 
are zero-based indices), but exponentiation is not needed to calculate 
it---only bitwise operations that directly set the corresponding bits. The 
amplitudes of the state vector $\psi$ are stored in an array with the 
positions corresponding to the binary numbers obtained by the above mapping. 
For the density matrix $\rho$, both the row and the column of an element are 
determined by this mapping. Thus, given a basis state specification (i.e., 
which spins are up and which are down), it allows to quickly determine the 
position of the corresponding state vector amplitude or density matrix element 
in the stored arrays. This leads to an efficient calculation of the action of 
operators both in the Schr\"odinger equation~\eqref{schroedinger_equation} and 
the master equation~\eqref{master_equation}.

The commutator term on the right-hand side of master 
equation~\eqref{master_equation} can be written
\begin{gather}\label{Hrho_term}
-\frac{i}{\hbar}[\hat{H},\rho]
=-\frac{i}{\hbar}\hat{H}\rho+\left(-\frac{i}{\hbar} \hat{H}\rho\right)^\dagger,
\end{gather}
where we have used the fact $\hat{H}$ and $\rho$ are Hermitian, i.e., 
$\hat{H}^\dagger=\hat{H}$ and $\rho^\dagger=\rho$. Writing the commutator term 
like this, shows that only $-\frac{i}{\hbar}\hat{H}\rho$ needs to be 
calculated by applying $-\frac{i}{\hbar}\hat{H}$ to each column of $\rho$ just 
like in a Schr\"odinger equation~\eqref{schroedinger_equation}, and then the 
adjoint of the resulting matrix can be added to itself to obtain 
$-\frac{i}{\hbar}[\hat{H},\rho]$. Hence, we first need to be able to calculate 
the action of $\hat{H}$ on a state vector $\psi$.

The operators $-\sum_j\Delta\hat{n}_{j}$, 
$\sum_{j\neq k} J_{z,jk} \hat{n}_{j}\hat{n}_{k}$ in the 
Hamiltonian~\eqref{H_long_range_ising_Rydberg}; and 
$-\sum_j\frac{\Delta_j}{2}\hat{\sigma}_{z,j}$, 
$\sum_{j\neq k} J_{z,jk} \hat{\sigma}_{z,j}\hat{\sigma}_{z,k}$ in the 
Hamiltonian~\eqref{H_long_range_ising} are diagonal. Therefore, their storage 
requirements are the same as for the state vector $\psi$ itself. Even though 
the action of these operators could be calculated on the fly, we find that for 
all of them except $-\sum_j\Delta\hat{n}_{j}$, storing the diagonal of the 
matrix and multiplying it onto $\psi$ is faster. The operator 
$-\sum_j\Delta\hat{n}_{j}$ has all the detunings being equal and can be 
calculated efficiently without any storage. It will be explained below, as the 
same approach is used for the homogeneous part of the operator 
$-\sum_j\frac{\Delta_j}{2}\hat{\sigma}_{z,j}$. In principle, however, action 
of all of these operators could be calculated in a completely matrix-free way 
by adapting the code below to immediately multiply each element of $\psi$ with 
a corresponding element of the operator diagonal instead of storing it.

\begin{table}[htb]
\centering
\begin{tabular}{|c | l|} 
\hline
Operator  &  Description \\
\hline
\verb+<<+  & bitwise shift\\
\verb+&+   & bitwise AND\\
\verb+|+   & bitwise OR\\
\verb+^+   & bitwise XOR\\
\verb+~+   & bitwise NOT\\
\verb+&&+  & logical AND\\
\verb+||+  & logical OR\\
\verb+!+   & logical NOT\\
\hline
\end{tabular}
\caption{Logical and bitwise operators used in the pseudo-code examples.}
\label{table_operators}
\end{table}

We will explain both this and the other computations in terms of the 
pseudo-code where the bitwise and logical operators follow the conventions of 
C and C++ languages. The operators are listed in Table~\ref{table_operators}. 
The loops in the pseudo-code are written in terms of zero-based indices. E.g., 
\verb+n+ is an index of the basis for which it holds that 
$0\leq \verb+n+ \leq 2^N-1$; and \verb+j+,\verb+k+ are indices of the spins 
for which it holds that $0\leq \verb+j+,\verb+k+ \leq N - 1$. The basis size 
is written compactly using the bitwise shift operator as \verb+1 << N+, which 
is equal to $2^N$. The variables are set using~\verb+=+, so that 
\verb.n = n + 1. means that $1$ is added to \verb+n+.

Assume that the detunings $\Delta_j$ are stored in an array \verb+Delta+, with 
the elements $\verb+Delta[j]+=\Delta_j$. To calculate the diagonal \verb+d+ 
(an array of size $2^N$ with elements \verb+d[n]+) of the operator 
$-\sum_j\frac{\Delta_j}{2}\hat{\sigma}_{z,j}$, the following pseudo-code could 
be used:
\begin{verbatim}
for n = 0 to (1 << N) - 1
    d[n] = 0
    for j = 0 to N - 1
        if n & (1 << j) then
            d[n] = d[n] - 0.5*Delta[j]
        else
            d[n] = d[n] + 0.5*Delta[j]
        end if
    end for
end for
\end{verbatim}
In the above, the diagonal element \verb+d[n]+ is initialized with the value 
zero. Then, iterating over the spin index \verb+j+, the 
expression \verb+n & (1 << j)+ is used to test whether bit \verb+j+ is set in 
the basis state index \verb+n+. Depending on the result of this test, 
\verb+0.5*Delta[j]+ is either subtracted from or added to \verb+d[n]+.

To add the contribution of the term
$\sum_{j\neq k} J_{z,jk} \hat{\sigma}_{z,j}\hat{\sigma}_{z,k}$ to the diagonal \verb+d+,
assume that the shifts $J_{z,jk}$ are stored 
in a matrix \verb+J_z+ with the elements $\verb+J_z[j][k]+=J_{z,jk}$ if 
$j\neq k$ and $\verb+J_z[j][k]+=0$ if $j=k$.
Then the following pseudo-code could be used:
\begin{verbatim}
for n = 0 to (1 << N) - 1
    for j = 0 to N - 1
        for k = 0 to N - 1
            nj = n & (1 << j)
            nk = n & (1 << k)
            if (nj && nk) || (!nj && !nk)
            then
                d[n] = d[n] + J_z[j][k]
            else
                d[n] = d[n] - J_z[j][k]
            end if
        end for
    end for
end for
\end{verbatim}
In the above, \verb+(nj && nk) || (!nj && !nk)+ evaluates to logical TRUE if 
both bits \verb+j+ and \verb+k+ are set or both bits are reset. In this case 
\verb+J_z[j][k]+ is added to \verb+d[n]+. Otherwise, \verb+J_z[j][k]+ is 
subtracted. The above code can be adjusted to realize the operator 
$\sum_{j\neq k} J_{z,jk} \hat{n}_{j}\hat{n}_{k}$ instead by using
\begin{verbatim}
            if (nj && nk)
            then
                d[n] = d[n] + J_z[j][k]
            end if
\end{verbatim}
in the inner loop, since $\hat{n}_{j}\hat{n}_{k}$ only gives a contribution
when both spins are excited.

In the Hamiltonian~\eqref{H_long_range_ising}, the detunings are written 
$\Delta_j=\Delta_\text{avg}+\omega_{{01},\text{avg}}-\omega_{01,j}$, where the 
detuning from the average frequency $\Delta_\text{avg}$ is also being 
optimized together with the pulse parameters $a_p$ and $b_p$. Thus, to avoid 
recomputation of the diagonal \verb+d+ for every new value of 
$\Delta_\text{avg}$, we store only $\omega_{{01},\text{avg}}-\omega_{01,j}$ in 
the array \verb+Delta+. The operator 
$-\frac{\Delta_\text{avg}}{2}\sum_j\hat{\sigma}_{z,j}$ is applied separately 
without additional storage. This is accomplished with the instruction 
\verb+popcount+ (population count) that counts the number of set bits in a 
binary number. Assuming that the variable \verb+Delta_avg+ holds the value of 
$\Delta_\text{avg}$, the state vector $\psi$ is stored in the array 
\verb+psi+, and the result $r=-\sum_j\frac{\Delta_j}{2}\hat{\sigma}_{z,j}\psi$ 
is stored in the array \verb+res+, the following pseudo-code can be used:
\begin{verbatim}
for n = 0 to (1 << N) - 1
    s = -0.5*Delta_avg*(2*popcount(n) - N)
    res[n] = (s + d[n])*psi[n]
end for
\end{verbatim}
The same approach is used to calculate the action of the operator 
$-\sum_j\Delta\hat{n}_{j}$ in the 
Hamiltonian~\eqref{H_long_range_ising_Rydberg} by changing
\begin{verbatim}
    s = -Delta*popcount(n)
\end{verbatim}
in the code above.

The rest of the operators in the 
Hamiltonians~\eqref{H_long_range_ising_Rydberg}~and~\eqref{H_long_range_ising} 
are non-diagonal. The driving term $-\Omega\sum_j\hat{\sigma}_{x,j}$ in the 
Hamiltonian~\eqref{H_long_range_ising_Rydberg} is the simplest to calculate. 
The action of this operator on the state vector $\psi$ can be implemented with 
the following pseudo-code:
\begin{verbatim}
for n = 0 to (1 << N) - 1
    sum = 0
    for j = 0 to N - 1
        sum = sum + psi[n ^ (1 << j)]
    end for
    res[n] = res[n] - Omega * sum
end for
\end{verbatim}
Applying the XOR operator in a loop over all the spin indices 
finds the matrix elements between the states that differ by one flipped 
spin.

The implementation of the driving terms
$-\sum_j\left(\Omega\hat{\sigma}_{+,j}
+\Omega^*\hat{\sigma}_{-,j}\right)$
in the Hamiltonian~\eqref{H_long_range_ising} is more involved. It can be done 
by iterating over the set or reset bits in a binary number. Modern CPUs 
provide fast instructions to accomplish this. One such instruction is 
\verb+ctz+ (count trailing zeros). Counting the number of the trailing zeros 
is the same as determining the position of the last set bit. Once the position 
of the last set bit in a binary number \verb+b+ is determined, it can be reset 
by assigning \verb+b = b & (b - 1)+. This can also be reduced to a single 
instruction (\verb+blsr+) on modern CPUs. By repeatedly finding and then 
resetting the last set bit, it is possible to efficiently iterate over all the 
set bits in the binary number \verb+b+.

The instructions \verb+popcount+ and \verb+ctz+ can be used in some C/C++ 
compilers with \verb+__builtin_popcount+ (or \verb+std::popcount+) and 
\verb+__builtin_ctz+ (or \verb+std::countr_zero+), respectively. This keeps 
the code portable, as the compiler will either use the native instructions or 
generate (slower) code that accomplishes the same operation, depending on the 
particular CPU.

In the operator $-\sum_jf_j\Omega\hat{\sigma}_{+,j}$, the factors $f_j$ are 
stored in the array \verb+OmegaFactors+. Then the action of this operator on 
the state vector $\psi$ can be implemented with the following pseudo-code:
\begin{verbatim}
for n = 0 to (1 << N) - 1
    sum1 = 0
    b = n
    while b != 0
        sb = ctz(b)
        f = OmegaFactors[sb]
        sum1 = sum1 + f*psi[n & ~(1 << sb)]
        b = b & (b - 1)
    end while
    res[n] = res[n] - Omega * sum1
end for
\end{verbatim}
Iterating over the set bits of \verb+n+, the expression \verb+n & ~(1 << sb)+ 
resets the bit at the position \verb+sb+ (the currently found set bit in 
\verb+n+), and the element of \verb+psi+ that has this new index is added to 
\verb+sum1+ with a weight \verb+OmegaFactors[sb]+. This sum is multiplied by 
the variable \verb+Omega+ that holds the value of $\Omega$.

Adding the contribution of the operator $-\sum_j f_j\Omega^*\hat{\sigma}_{-,j}$ 
can be implemented with
\begin{verbatim}
for n = 0 to (1 << N) - 1
    sum2 = 0
    b = ~n & ((1 << N) - 1)
    while b != 0
        sb = ctz(b)
        f = OmegaFactors[sb]
        sum2 = sum2 + f*psi[n | (1 << sb)]
        b = b & (b - 1)
    end while
    res[n] = res[n] - conj(Omega) * sum2
end for
\end{verbatim}

Instead of iterating over the set bits, this loop iterates over the reset 
bits, by using the bitwise NOT to transform the binary number \verb+b+ 
compared to the previous loop. When using fixed-size integers for \verb+n+ and 
\verb+b+, the bits above the maximum number of spins are unphysical. However, 
\verb+~n+ has these bits set, and to prevent them from interfering with the 
subsequent loop that should only iterate over the physically meaningful bits, 
they are reset by performing the bitwise AND with \verb+(1 << N) - 1+. Inside 
the loop, the expression \verb+n | (1 << sb)+ sets the bit at the position 
\verb+sb+, and the element of \verb+psi+ that has this new index is added to 
\verb+sum2+ with a weight \verb+OmegaFactors[sb]+. This sum is multiplied by 
\verb+conj(Omega)+ that holds the value of $\Omega^*$.

The term $\sum_{j\neq k} J_{\pm,jk} \hat{\sigma}_{+,j}\hat{\sigma}_{-,k}$ in 
the Hamiltonian~\eqref{H_long_range_ising} can also be implemented by 
iterating over the set and reset bits. Assuming that the matrix \verb+J_pm+ 
has elements $\verb+J_pm[j][k]+=J_{\pm,jk}$, the following pseudo-code can be 
used:
\begin{verbatim}
for n = 0 to (1 << N) - 1
    sum = 0
    b = ~n & ((1 << N) - 1)
    while b != 0
        sb = ctz(b)
        c = n
        while c != 0
            sc = ctz(c)
            ns = (n & ~(1 << sc)) | (1 << sb)
            sum = sum + J_pm[sc][sb]*psi[ns]
            c = c & (c - 1)
        end while
        b = b & (b - 1)
    end while
    res[n] = res[n] + sum
end for
\end{verbatim}

The code above describes the reduced-storage multiplication of a Hamiltonian 
$\hat{H}$ onto a state vector $\psi$. For the master 
equation~\eqref{master_equation}, the code needs to operate on each column of 
the density matrix $\rho$. Additionally, the reduced-storage action of the 
dissipators is needed. In the sum of the terms 
$\gamma\sum_j\mathcal{D}[\hat{\sigma}_{-,j}]\rho
+\gamma_\text{d}\sum_j\mathcal{D}[\hat{\sigma}_{z,j}]\rho
$ in the master equation~\eqref{master_equation}, only the term 
$\sum_{j}\hat{\sigma}_{-,j}\rho\hat{\sigma}_{+,j}$ involves a non-diagonal 
superoperator acting on $\rho$. The diagonal terms can be written
\begin{gather}
\begin{aligned}
&\gamma\sum_j\mathcal{D}[\hat{\sigma}_{-,j}]\rho
+\gamma_\text{d}\sum_j\mathcal{D}[\hat{\sigma}_{z,j}]\rho-\gamma\sum_{j}\hat{\sigma}_{-,j}\rho\hat{\sigma}_{+,j}\\
&=-\frac{\gamma}{2}\sum_j(\rho\hat{\sigma}_{11,j}+\hat{\sigma}_{11,j}\rho)
-\gamma_\text{d}\sum_j(\rho-\hat{\sigma}_{z,j}\rho\hat{\sigma}_{z,j})\\
&= D \odot \rho,
\end{aligned}
\end{gather}
where $\odot$ is the elementwise (Hadamard) matrix product, 
$\hat{\sigma}_{11,j}=\hat{\sigma}_{+,j}\hat{\sigma}_{-,j}$, and we have used 
that $\hat{\sigma}_{z,j}\hat{\sigma}_{z,j}$ is the identity matrix. Since 
$\hat{\sigma}_{11,j}$ and $\hat{\sigma}_{z,j}$ are diagonal matrices, we 
denote their diagonal elements by $\sigma_{11,j,m}$ and $\sigma_{z,j,m}$, 
respectively. Then the elements of the matrix $D$ are
\begin{gather}
\begin{aligned}
&D_{mn}=-\frac{\gamma}{2}\sum_j(\hat{\sigma}_{11,j,m}+\hat{\sigma}_{11,j,n})\\
&-\gamma_\text{d}\left(N-\left(\sum_j \hat{\sigma}_{z,j,m}\hat{\sigma}_{z,j,n}\right)\right).
\end{aligned}
\end{gather}

The non-diagonal superoperator 
$\sum_{j}\hat{\sigma}_{-,j}\rho\hat{\sigma}_{+,j}$ can be realized by 
considering its elements
\begin{gather}\label{non_diagonal_superoperator}
\sum_{j}\langle m|\hat{\sigma}_{-,j}\rho\hat{\sigma}_{+,j}|n\rangle.
\end{gather}
This sum has non-zero terms when for each spin $j$, both of the basis states 
$|m\rangle$ and $|n\rangle$ have this spin in state $|0\rangle$. Assuming that 
the binary number \verb+m+ corresponds to the basis state $|m\rangle$, and the 
binary number \verb+n+ corresponds to the basis state $|n\rangle$, the 
non-zero elements of Eq.~\ref{non_diagonal_superoperator} can be found by 
iterating over the reset bits of the binary number \verb+m | n+. This is 
equivalent to iterating over the set bits of the binary number 
\verb+~m & ~n & ((1 << N) - 1)+, where we have used the equivalence
\begin{gather}
\text{NOT}(A\text{ OR }B)=(\text{NOT }A)\text{ AND }(\text{NOT }B)
\end{gather}
and masked off the unphysical bits by the bitwise AND with \verb+(1 << N) - 1+. 

If we define the variable \verb+gamma+ that has the value of $\gamma$, the 
matrix \verb+mul+ with the elements $\verb+mul[m][n]+=D_{mn}$, and the matrix 
\verb+rho+ that holds the elements of $\rho$, the result of 
the expression $\gamma\sum_j\mathcal{D}[\hat{\sigma}_{-,j}]\rho
+\gamma_\text{d}\sum_j\mathcal{D}[\hat{\sigma}_{z,j}]\rho$ 
added to the matrix \verb+res+ can be found using the following pseudo-code:
\begin{verbatim}
for n = 0 to (1 << N) - 1
    for m = 0 to n
        sum = 0
        b = ~m & ~n & ((1 << N) - 1)
        while b != 0
            sb = ctz(b)
            ms = m | (1 << sb)
            ns = n | (1 << sb)
            sum = sum + rho[ms][ns]
            b = b & (b - 1)
        end while
        res[m][n] = res[m][n]
                    + gamma*sum
                    + rho[m][n]*mul[m][n]
    end for
end for
\end{verbatim}

Since $\rho$ is a Hermitian matrix, its lower triangular part is not stored 
explicitly. Therefore, the variable \verb+m+ (row index) in the above code is 
iterated over from $0$ to \verb+n+ (instead of \verb+(1 << N) - 1+). Omitting 
the lower triangular part of $\rho$ does not introduce complexity in this 
code, since both the variables \verb+ms+ and \verb+ns+ are obtained from 
\verb+m+ and \verb+n+, respectively, by setting the same bit \verb+sb+. This 
effectively adds the same constant to both. Hence, it holds that 
$\verb+ms+\leq\verb+ns+$. However, special handling is required for the lower 
triangular part of $\rho$ when calculating $\hat{H}\rho$. There, $\rho$ gets 
temporarily uncompressed (filling the lower triangular part) and then 
compressed again when evaluating Eq.~\eqref{Hrho_term}.

\section{Calculation of the gradient}\label{app_gradient}
In this appendix, we describe the calculation of the gradient used in the 
optimal control. The approach is described in detail in the appendix of
Ref.~\cite{iakoupov_prresearch23}, and here we only discuss the differences. 
One of them is reducing the storage requirement of the operators and 
superoperators using App.~\ref{app_reduced_storage}. Another difference is 
that both the Schr\"odinger equation and master equation are used to evolve 
the initial states and optimize the pulse shapes. For both equations, the 
detuning $\Delta_\text{avg}$ from the average spin frequency is also being 
optimized together with the pulse shape parameters $a_p$ and $b_p$.

We discuss the master equation case first, as this is what 
Ref.~\cite{iakoupov_prresearch23} considered. In the analysis, the density 
matrix $\rho$ was rewritten as a vector $\vec{\rho}$, so that the master 
equation~\eqref{master_equation} is rewritten as 
\begin{gather}\label{master_equation_rho_vector}
\dot{\vec{\rho}}=\hat{L}\vec{\rho}, 
\end{gather}
where $\hat{L}$ is a matrix.
While such rewriting is useful conceptually, it can make the computations less 
efficient. Taken literally, it can, for instance, significantly increase the 
storage requirements. If $\vec{\rho}$ is in the row-major form 
($\rho_{l,l'}=(\vec{\rho})_{l2^N+l'}$), then $-\frac{i}{\hbar}[\hat{H},\rho]$ 
in the master equation~\eqref{master_equation} is rewritten as 
$-\frac{i}{\hbar}(\hat{H}\otimes \hat{I} - \hat{I}\otimes \hat{H}^T)\vec{\rho}$, 
where $\hat{I}$ is the identity operator, for the corresponding term in the 
vector master equation~\eqref{master_equation_rho_vector}. The tensor products  
$\hat{H}\otimes \hat{I}$ are $2^N$ copies of the Hamiltonian $\hat{H}$ and 
therefore have equal or larger storage requirements than the density matrix 
$\rho$. Since $\hat{H}$ is usually a sparse matrix, it has lower storage 
requirements than $\rho$, and calculating $-\frac{i}{\hbar}[\hat{H},\rho]$ 
through Eq.~\eqref{Hrho_term} avoids unnecessary storage. Additionally, as 
App.~\ref{app_reduced_storage} has shown, neither $\hat{H}$ nor the 
dissipators $\mathcal{D}$ need to be stored to find their action on $\rho$, 
but we store their diagonals as an optimization.

For the master equation, the calculation of the gradient of 
Eq.~\eqref{open_population_diff_def} reduces to calculation of the gradients 
\begin{gather}\label{open_P_gradient}
\begin{aligned}
\dpd{P}{a_p}=\vec{M}^\dagger\dpd{\vec{\rho}_{N_t}}{a_p},&&
\dpd{P}{b_p}=\vec{M}^\dagger\dpd{\vec{\rho}_{N_t}}{b_p},
\end{aligned}
\end{gather}
where $\vec{M}$ is the measurement operator $\hat{M}$ in 
Eq.~\eqref{open_population_diff_def} written as a vector, $N_t$ is the number 
of time steps, and $\vec{\rho}$ is defined on the discrete times $t_n=n(\Delta t)$ 
with $\Delta t=t_\text{f}/N_t$ such that $\vec{\rho}_n=\vec{\rho}(t_n)$.
The gradients in Ref.~\cite{iakoupov_prresearch23} are of the same form as 
Eqs.~\eqref{open_P_gradient}. The calculations are written in terms of 
$\hat{L}_{1,n}=\hat{L}(t_n)\Delta t$, 
$\hat{L}_{2,n}=\hat{L}(t_n+(\Delta t)/2)\Delta t$, 
$\hat{L}_{3,n}=\hat{L}(t_{n+1})\Delta t$,
\begin{subequations}
\begin{align}
&T_\text{Re}=-\frac{i}{\hbar}(\Delta t)
(\tilde{H}_\text{d,Re}\otimes \hat{I}
- \hat{I}\otimes \tilde{H}_\text{d,Re}^T),\\
&T_\text{Im}=-\frac{i}{\hbar}(\Delta t)
(\tilde{H}_\text{d,Im}\otimes \hat{I}
- \hat{I}\otimes \tilde{H}_\text{d,Im}^T),
\end{align}
\end{subequations}
where 
$\tilde{H}_\text{d,Re}=-\hbar\sum_j(\hat{\sigma}_{+,j}+\hat{\sigma}_{-,j})$ 
and 
$\tilde{H}_\text{d,Im}=-i\hbar\sum_j(\hat{\sigma}_{+,j}-\hat{\sigma}_{-,j})$ 
for the Hamiltonian~\eqref{H_long_range_ising}. As an extension, we also 
calculate $\pd{P}{\Delta_\text{avg}}$, for which, instead of $T_\text{Re}$ or 
$T_\text{Im}$,
\begin{align}
T_{\Delta_\text{avg}}=-\frac{i}{\hbar}(\Delta t)
(\tilde{H}_{\Delta_\text{avg}}\otimes \hat{I}
- \hat{I}\otimes \tilde{H}_{\Delta_\text{avg}}^T)
\end{align}
is used with $\tilde{H}_{\Delta_\text{avg}}=-(\hbar/2)\sum_j\hat{\sigma}_{z,j}$. 
The action of $T_\text{Re}$, $T_\text{Im}$, $T_{\Delta_\text{avg}}$, and 
$\hat{L}$ on $\vec{\rho}$ is calculated using Eq.~\eqref{Hrho_term} instead of 
storing the tensor product matrices.

The Schr\"odinger equation is of the same form as 
Eq.~\eqref{master_equation_rho_vector} with the replacements 
$\vec{\rho}\rightarrow \psi$ and $\hat{L}\rightarrow -i\hat{H}/\hbar$. The 
expression for $P_1-P_0$ has a slightly different form, however. The spin 
amplification superoperator $\mathcal{U}$ in 
Eq.~\eqref{open_population_diff_def} can be written as 
$\mathcal{U}(\rho)=\hat{U}\rho \hat{U}^\dagger$ with unitary operator $\hat{U}$. 
Hence, Eq.~\eqref{open_population_diff_def} becomes
\begin{gather}\label{closed_population_diff_def}
P_1-P_0=
\psi_1^\dagger\hat{U}^\dagger \hat{M} \hat{U}\psi_1
-\psi_0^\dagger\hat{U}^\dagger \hat{M} \hat{U}\psi_0.
\end{gather}
Instead of Eqs.~\eqref{open_P_gradient}, the gradient is calculated by 
evaluating
\begin{subequations}
\begin{gather}\label{closed_P_gradient}
\dpd{P}{a_p}
=2\text{Re}\left[\psi_{N_t}^\dagger \hat{M} \dpd{\psi_{N_t}}{a_p}\right],\\
\dpd{P}{b_p}
=2\text{Re}\left[\psi_{N_t}^\dagger \hat{M} \dpd{\psi_{N_t}}{b_p}\right].
\end{gather}
\end{subequations}
The subexpressions
\begin{gather}\label{closed_P_gradient_sub}
\begin{aligned}
\psi_{N_t}^\dagger \hat{M} \dpd{\psi_{N_t}}{a_p},
&& \psi_{N_t}^\dagger \hat{M} \dpd{\psi_{N_t}}{b_p},
\end{aligned}
\end{gather}
are of the same form as Eqs.~\eqref{open_P_gradient}, and hence can be 
calculated in the similar way. Setting 
$\hat{L}_{1,n}=-i\hat{H}(t_n)\Delta t/\hbar$, 
$\hat{L}_{2,n}=-i\hat{H}(t_n+(\Delta t)/2)\Delta t/\hbar$, 
$\hat{L}_{3,n}=-iH(t_{n+1})\Delta t/\hbar$, 
\begin{subequations}
\begin{align}
&T_\text{Re}=-\frac{i}{\hbar}(\Delta t)\tilde{H}_\text{d,Re},\\
&T_\text{Im}=-\frac{i}{\hbar}(\Delta t)\tilde{H}_\text{d,Im},
\end{align}
\end{subequations}
makes is possible to reuse the derivations of 
Ref.~\cite{iakoupov_prresearch23} to evaluate 
Eqs.~\eqref{closed_P_gradient_sub}. Similar to the master equation, 
$\pd{P}{\Delta_\text{avg}}$ can be calculated by replacing $T_\text{Re}$ or 
$T_\text{Im}$ with
\begin{align}
T_{\Delta_\text{avg}}=-\frac{i}{\hbar}(\Delta t)\tilde{H}_{\Delta_\text{avg}}.
\end{align}

\section{Continuous-wave drive spin amplification in realistic systems}\label{app_cw_drive}

In this appendix, we show that the imperfections commonly found in realistic 
spin ensembles, make the weak continuous-wave pulses used in 
Refs.~\cite{lee_pra05,furman_prb06,close_prl11} perform suboptimally. At the 
end of this appendix, we also consider the frequency-ramp pulse that is known 
to perform well for the adiabatic state preparation~\cite{omran_science19}, 
but applied to the spin amplification, we found it to only work well for the 
amplification times that are much longer than the coherence times in the 
realistic spins. 

To show the effect of each imperfection separately, the detailed models for 
the realistic spins are not used in this appendix.  The ensemble is chosen to 
be a 1D array of $16$ spins. For the idealized nearest-neighbor 
transverse-field Ising model, we use the 
Hamiltonian~\eqref{H_long_range_ising} with $J_{\pm,jk}=0$, $J_{z,jk}=J_z$, 
$J_z>0$ when $j$ and $k$ are nearest neighbors, $J_{z,jk}=0$ otherwise. The 
drive has a constant Rabi frequency $\Omega/J_z=0.1$ with $f_j=1$ and 
detunings $\Delta_j=0$, where all spins have the same transition frequency.  
Unless noted otherwise, the localized initial state 
$|\psi_1\rangle = \hat{\sigma}_{+,j=1}|\psi_0\rangle$ will be used in this 
appendix, with $j=1$ being the end spin in 1D~\cite{lee_pra05} and corner spin 
in 2D~\cite{close_prl11}.

In 1D, the simple explanation of the spin amplification is as 
follows~\cite{lee_pra05,furman_prb06,close_prl11}. The interaction term 
$J_z \hat{\sigma}_{z,j}\hat{\sigma}_{z,k}$ (for the nearest-neighbor $j$ and 
$k$) in the Hamiltonian~\eqref{H_long_range_ising} causes effective detunings. 
When each spin is in the ground state $|0_j\rangle$, the drive is effectively 
off-resonant with the detuning $8J_{z}$ in the bulk and with $4J_{z}$ for the 
end spins. Note that there are factors of $2$ here compared to 
Ref.~\cite{close_prl11}. One factor of $2$ is from the 
term $(\Delta_j/2)\sigma_{z,j}$, and another one is because the summation over 
$J_{z,jk} \hat{\sigma}_{z,j}\hat{\sigma}_{z,k}$ in the 
Hamiltonian~\eqref{H_long_range_ising} double counts the interaction terms 
with $j>k$ and $j<k$. As soon as one spin is excited to the state $|1_j\rangle$, 
the detuning on the neighboring spins due to the interaction term vanishes, 
and hence a resonant drive can create further excitations. The additional 
excitations are created due to a propagating domain wall around the initial 
one.

\begin{figure}[t]
\begin{center}
\includegraphics{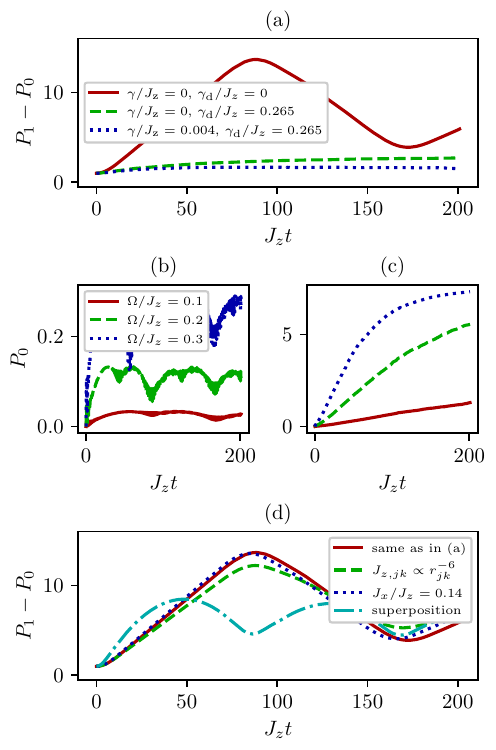}
\end{center}
\caption{(a) The differences of the excited spin populations $P_1$ and $P_0$ 
as functions of time. The 1D ensemble of $16$ spins is described by the 
Hamiltonian~\eqref{H_long_range_ising} with $J_{\pm,jk}=0$, nearest-neighbor 
$J_{z,jk}=J_z$, $J_z>0$, $\Omega/J_z=0.1$ with $f_j=1$, and $\Delta_j=0$. For 
$P_1$, the initial state is with a single excitation in the end spin. (b,c) 
Populations $P_0$ with the initial zero-excitation state as functions of time. 
The parameters are the same as in (a), except for $\Omega$. (b) Using
$\gamma_\text{d}=\gamma=0$. Rapid oscillations make the curves appear to 
change thickness. (c) Using $\gamma_\text{d}/J_z=0.265$ and $\gamma/J_z=0.004$. 
The curves with the non-zero $\gamma_\text{d}$ or $\gamma$ have discontinuities 
due to using the quantum trajectory method with $500$~trajectories instead of 
the master equation for the numerical calculations. (d) Using 
$\gamma_\text{d}=\gamma=0$, but where the Hamiltonian is modified to either 
have a long-range $J_{z,jk}=J_z r_{jk}^{-6}$ with $r_{jk}=|j-k|$ (dashed 
green) or $J_x/J_z = 0.14$ (dotted blue). The dash-dotted cyan curve shows the 
effect of using the equal superposition state~\eqref{psi_1_definition} as the 
initial one for $P_1$.
\label{fig_weak_drive}}
\end{figure}

This is reflected in the numerical simulations. As shown by the solid red 
curve in Fig.~\ref{fig_weak_drive}(a), putting a single excitation in the end 
spin, the number of excited spins increases. The amplification is stopped and 
then reversed due the propagating domain wall being reflected at the other end 
of the ensemble. The dashed green curve in Fig.~\ref{fig_weak_drive}(a) shows 
the effect of added pure dephasing, i.e., with $\gamma_\text{d}/J_z=0.265$ and 
$\gamma=0$ in the master equation~\eqref{master_equation} (evaluated using the 
quantum trajectory method~\cite{daley_tandf14}). The dotted blue curve in 
Fig.~\ref{fig_weak_drive}(a) uses $\gamma_\text{d}/J_z=0.265$ and 
$\gamma/J_z=0.004$, i.e., both pure dephasing and population decay. The decay 
rates are chosen to approximately match $T_1$ and $T_\phi$ used for the 
persistent-current artificial atoms in the main text. Assuming
$J_z=2\pi\cdot40$~MHz, $T_1=1\text{ }\mu\text{s}$, and 
$T_{\phi}=15\text{ }\text{ns}$, we have $\gamma/J_z=1/(T_1J_z)\approx 0.004$ 
and $\gamma_\text{d}/J_z=1/(2T_{\phi}J_z)\approx 0.265$.

For the initial zero-excitation state, using $\gamma_\text{d}=\gamma=0$ 
results in off-resonant Rabi oscillations where all the spins precess close to 
the state $|0_j\rangle$. This is shown in Fig.~\ref{fig_weak_drive}(b), where 
the additional complexity in the dynamics stems from including the spin-spin 
interactions. Importantly, the total excitation number is bounded for a given 
constant $\Omega$. When the dephasing is added, it can be viewed as a 
continuous measurement of the spin state by the environment, and therefore 
even a small amplitude of the state~$|1_j\rangle$ can lead to a steady 
accumulation of the number of excited spins. This accumulation happens faster 
for stronger drives, as they make the spins acquire a larger amplitude of the 
state~$|1_j\rangle$. This is shown in Fig.~\ref{fig_weak_drive}(c) that was 
obtained using $\gamma_\text{d}/J_z=0.265$ and $\gamma/J_z=0.004$. In the 
contrast to the $\gamma_\text{d}=\gamma=0$ case, the populations are 
increasing over time. Despite this, a larger $\Omega$ can improve $P_1-P_0$. 
Doing the sweep over $\Omega \in [0.2, 2.4]\cdot J_z$ and 
$\Delta_\text{avg}\in[-5,5]\cdot J_z$, we find maximum $P_1-P_0=2.49$ 
for $\Omega/J_z=0.4$ and $\Delta_\text{avg}=0$. This is slightly better than 
$P_1-P_0=1.69$ for $\Omega/J_z=0.1$ (maximum of the dotted blue curve in 
Fig.~\ref{fig_weak_drive}(a)).

\begin{figure}[t]
\begin{center}
\includegraphics{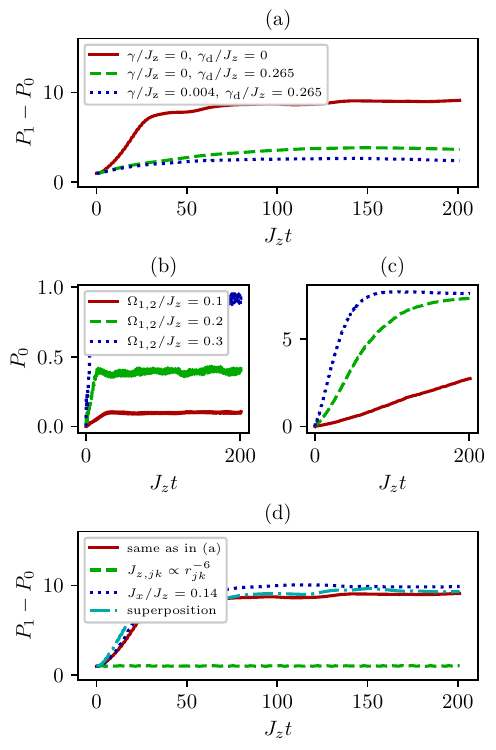}
\end{center}
\caption{This is the equivalent of Fig.~\ref{fig_weak_drive}, but where the 
$16$ spins are arranged in a 2D square lattice instead of a 1D chain. (a),(d) 
The differences of the excited spin populations $P_1$ and $P_0$ as functions 
of time. (b),(c) Populations $P_0$ with the initial zero-excitation state as 
functions of time. A two-frequency drive is used with the
parametrization~\eqref{two_frequency_Omega}. The parameters are 
$\Omega_{1}/J_z=\Omega_{2}/J_z=0.1$ (for (a,d)) $\Delta_{\text{avg},1}=0$ and 
$\Delta_{\text{avg},2}/J_z=-4$.
\label{fig_2d_weak_drive}}
\end{figure}

Even without decay and decoherence, deviations of the 
Hamiltonian~\eqref{H_long_range_ising} from the nearest-neighbor Ising model 
also reduce the quality of the spin amplification. This is illustrated in 
Fig.~\ref{fig_weak_drive}(d). The long-range interaction can be detrimental if 
the next-nearest-neighbor shift is larger than $\Omega$. This puts a 
limitation on how weak $\Omega$ can be in practice. Assuming 
$J_{z,jk}=J_z r_{jk}^{-6}$, the next-nearest-neighbor shift in 1D is 
$4|J_{z,jk}|/J_z=1/16$ (with $r_{jk}=|j-k|$), and it is smaller than the 
chosen $\Omega/J_z=0.1$, so that spin amplification still works, as shown by 
the dashed green curve in Fig.~\ref{fig_weak_drive}(d). However, for 
$J_{z,jk}=J_z r_{jk}^{-3}$, the next-nearest-neighbor shift is 
$4|J_{z,jk}|/J_z=1/2$, which is larger than $\Omega/J_z=0.1$, and this 
completely shuts down spin amplification. We do not show the latter case in 
Fig.~\ref{fig_weak_drive}(d), but it looks very similar to the dashed green 
curve in Fig.~\ref{fig_2d_weak_drive}(d) that shows $J_{z,jk}=J_z r_{jk}^{-6}$ 
in 2D and will be discussed below.

The long-range coupling $J_{z,jk}=J_z r_{jk}^{-3}$ is the asymptotic form of 
the dipole-dipole interaction of the persistent-current artificial atoms, but 
it has significant corrections when they are close to each other. It can be 
analytically evaluated using the Neumann formula for the mutual inductance. 
For the Rydberg atoms, the coupling $J_{z,jk}=J_z r_{jk}^{-6}$ is more 
natural. Regardless of the power in the long-range interaction, it is possible 
to achieve some spin amplification by choosing a larger
$\Omega$~\cite{furman_prb06} to compensate for the next-nearest-neighbor shifts 
and beyond. As discussed above, this will need to be balanced with the other 
imperfections that become more significant for larger $\Omega$.

Another imperfection of the Hamiltonian shown in Fig.~\ref{fig_weak_drive}(d) 
is the non-zero $J_x/J_z = 0.14$ (dotted blue), which is approximately equal 
to the ratio calculated from the full model of the persistent-current 
artificial atoms for the chosen parameters. This also leads to a minor 
decrease in spin amplification. Finally, even if the Hamiltonian is the ideal 
nearest-neighbor Ising model, but the initial single-excitation state is a 
superposition of all the possible positions in the ensemble instead of being 
localized in the end spin, the continuous-wave drive spin amplification has a 
significantly lower maximum $P_1-P_0$ (dash-dotted cyan).

The 2D version of the continuous-wave spin amplification is illustrated in 
Fig.~\ref{fig_2d_weak_drive}. The intuitive explanation~\cite{close_prl11} 
builds upon the one from the 1D case (described in the beginning of this 
appendix). Again accounting for factors of $2$ resulting from the different 
scaling of the $\Delta_j$ terms and double counting of the $J_{z,jk}$ terms 
compared to Ref.~\cite{close_prl11}, the argument is that in 2D, when all 
spins are in the ground state, the corner spins are effectively detuned by 
$8J_{z}$, the edge spins by $12J_{z}$, and the bulk spins by $16J_{z}$. This 
is due to the different number of the nearest neighbors that the spins in the 
different locations have. Setting the corner spin to the excited state makes 
the neighboring edge spins be detuned by only $4J_{z}$. Thus, introducing 
another drive with $\Delta_{\text{avg},2}=-4J_{z}$ makes the edge spins become 
excited. Once they are, the neighboring bulk spins see their detuning 
effectively vanish, and hence the same resonant drive as for the 1D case can 
excite them. Thus, a two-frequency continuous-wave drive is sufficient to 
perform spin amplification in an idealized 2D ensemble.

\begin{figure}[t]
\begin{center}
\includegraphics{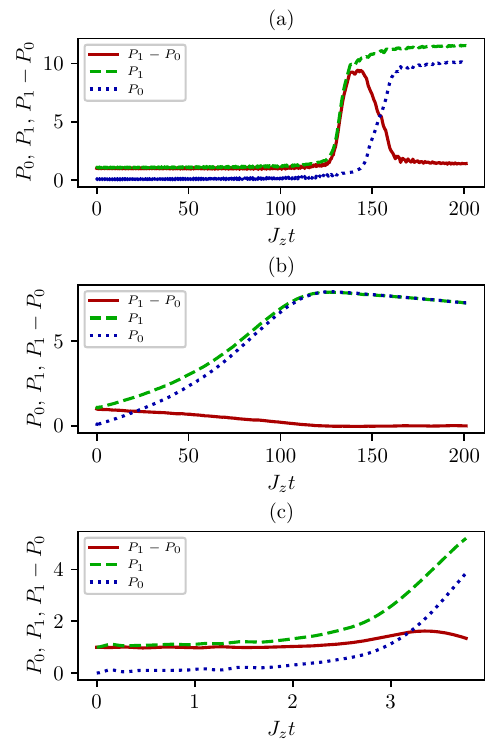}
\end{center}
\caption{Frequency ramp protocol where $\Delta$ is changed linearly from 
$\Delta/J_z=10$ to $\Delta/J_z=-10$ in the shown time interval. In all 
subplots, $16$ spins are arranged in a 2D square lattice with 
$J_{z,jk}=J_z r_{jk}^{-6}$, $\Omega/J_z=1$. (a) $\gamma/J_{\rm z}=0$, 
$\gamma_{\rm d}/J_z=0$ (b) $\gamma/J_{\rm z}=0.004$, $\gamma_{\rm d}/J_z=0.265$ 
(c) Short amplification time with $\gamma/J_{\rm z}=0.004$, and 
$\gamma_{\rm d}/J_z=0.265$.
\label{fig_Delta_ramp}}
\end{figure}

There are some qualitative differences compared to the analysis of 
Ref.~\cite{close_prl11}. This is due to the fact that we simulate the full 
nearest-neighbor transverse-field Ising model with the exponential basis size 
of the Hilbert space instead of approximating it by an effective tunneling 
model with a reduced basis~\cite{close_prl11}. Because the ensemble has finite 
size, boundary effects make the amplification stop at around half of the 
maximal value $P_1-P_0=16$. The achieved amplification stays at the same level 
in contrast with the 1D case where the total population starts decreasing 
again after the domain wall interacts with the boundary [cf. 
Fig.~\ref{fig_weak_drive}(a)]. There is also a more subtle difference in that 
the population of the initial corner spin decreases to around the same level 
as the steady state of the other spins in the ensemble. This is because we do 
not assume it to be of a different species with a different transition 
frequency like in Ref.~\cite{close_prl11}. The second drive frequency appears 
to be responsible for this population decrease, because it does not happen in 
1D with a single-frequency drive.

The coherent deviations from the idealized model also have different behavior 
than in 1D [Fig.~\ref{fig_2d_weak_drive}(d) compared to 
Fig.~\ref{fig_weak_drive}(d)]. Using the superposition state with the 
nearest-neighbor Hamiltonian has less difference in 2D (dash-dotted cyan 
curves). The 2D setup is more sensitive to the long-range interactions, 
however. This is because the next-nearest neighbors are closer in 2D than in 
1D (across the diagonal instead of twice the distance in the same direction). 
Hence, even with $J_{z,jk}=J_z r_{jk}^{-6}$, the next-nearest-neighbor shift 
is $4|J_{z,jk}|/J_z=1/2$, which is exactly the same as the 
next-nearest-neighbor shift for 1D with $J_{z,jk}=J_z r_{jk}^{-3}$. This shift 
is larger than $\Omega/J_z=0.1$, and this prevents spin amplification from 
working.

A different pulse type, which was used for the adiabatic state preparation in 
Ref.~\cite{omran_science19}, could also be applied here. In 
Fig.~\ref{fig_Delta_ramp}, $\Delta$ is changed linearly from $\Delta/J_z=10$ 
to $\Delta/J_z=-10$ over the shown time interval. The 2D ensemble with 
$J_{z,jk}=J_z r_{jk}^{-6}$ is used to illustrate that this pulse type can 
handle the long-range interactions. The case with $\gamma_{\rm d}=\gamma=0$ is 
shown in Fig.~\ref{fig_Delta_ramp}(a) and results in maximal $P_1-P_0=9.4$. 
However, once finite coherence time is added with $\gamma/J_{\rm z}=0.004$ and 
$\gamma_{\rm d}/J_z=0.265$, the difference $P_1-P_0$ does not rise above the 
initial value of unity, as shown in Fig.~\ref{fig_Delta_ramp}(b). This is 
because the $\Delta$ ramp time here is much larger than 
$1/\gamma_{\rm d}=3.77/J_z$. Changing $\Delta$ over this shorter time, does 
not recover any meaningful $P_1-P_0$, as shown in 
Fig.~\ref{fig_Delta_ramp}(c).

The failure of the $\Delta$-ramp pulse can be understood from the fact that 
spin amplification is not a single-state transfer protocol. State transfer of 
any of the two initial states $|\psi_0\rangle$ and $|\psi_1\rangle$ 
(Eqs.~\eqref{psi_0_definition}~and~\eqref{psi_1_definition}) is guaranteed by 
the adiabatic theorem and is robust to model imperfections, but spin 
amplification appears as a nonadiabatic effect where the $\Delta$ ramp needs 
to be stopped precisely when $|\psi_1\rangle$ is transferred to a 
highly-excited state, but $|\psi_0\rangle$ is not, as 
Fig.~\ref{fig_Delta_ramp}(a) shows. For the realistic spins discussed in the 
main text, such a window of $\Delta$ values is narrow for the Rydberg atoms 
and does not exist at all for the persistent-current artificial atoms due to 
the strong dephasing and large inhomogeneity. The latter is illustrated by 
Figs.~\ref{fig_Delta_ramp}(b)~and~\ref{fig_Delta_ramp}(c), albeit with a 
more idealized model than used in the main text for the persistent-current 
artificial atoms.

\bibliography{references}
\end{document}